\documentclass[fleqn,usenatbib]{mnras}

\usepackage[T1]{fontenc}
\usepackage{ae,aecompl}
\usepackage{csquotes}
\usepackage{threeparttable}

\DeclareRobustCommand{\VAN}[3]{#2}
\let\VANthebibliography\thebibliography
\def\thebibliography{\DeclareRobustCommand{\VAN}[3]{##3}\VANthebibliography}

\usepackage{mathtools}
\usepackage[varg]{txfonts}
\usepackage{color}
\usepackage{natbib}
\usepackage{multirow}   
\usepackage{gensymb}    
\usepackage{graphicx}
\graphicspath{ {Figures/} }
\usepackage{url}
\usepackage{indentfirst}
\usepackage[normalem]{ulem}
\usepackage{placeins}
\usepackage{tablefootnote}
\usepackage{amsmath}	
\usepackage{amssymb}	
\usepackage{mwe}
\hypersetup{colorlinks=true,allcolors=blue}
\newcommand{\bh}{4U~1543-47}

\defcitealias{nustar_paper2022}{DR22}
\title[Evolution of \bh\ during its 2021 outburst]{Probing the soft state evolution of \bh\ during its 2021 outburst using AstroSat}

\author[Husain et al. 2023]{
Nazma Husain$^{1}$,
Yash Bhargava$^{2,3}$,
Akash Garg$^{3}$,
Sneha Prakash Mudambi$^{4}$,
Ranjeev Misra$^{3}$,
Somasri Sen$^{1}$
\\
$^{1}$Department of Physics, Jamia Millia Islamia, New Delhi-110025, India\\
$^{2}$Tata Institute of Fundamental Research, Mumbai-400005, India \\
$^{3}$Inter-University Centre for Astronomy and Astrophysics, Pune-411007, India \\
$^{4}$CHRIST (Deemed to be University), Bengaluru-560029, India \\
}

\date{Accepted XXX. Received YYY; in original form ZZZ}

\pubyear{2023}
\begin{document}
\label{firstpage}
\maketitle


\begin{abstract}

\bh\ underwent its brightest outburst in 2021 after two decades of inactivity. During its decay phase, AstroSat conducted nine observations of the source spanning from July $1^{st}$ to September $26^{th}$, 2021. The first three observations were performed with an offset of 40$\arcmin$ with AstroSat/LAXPC, while the remaining six were on-axis observations. In this report, we present a comprehensive spectral analysis of the source as it was in the High/Soft state during the entire observation period. The source exhibited a disk-dominated spectra with a weak high-energy tail (power-law index $\geq2.5$) and a high inner disk temperature ($\sim$0.84~keV). Modelling the disk continuum with non-relativistic and relativistic models, we find inner radius to be significantly truncated at $>$$10~R_g$. Alternatively, to model the spectral evolution with the assumption that the inner disk is at the ISCO, it is necessary to introduce variation in the spectral hardening in the range $\sim$1.5-1.9.

\end{abstract}


\begin{keywords}
accretion: accretion disks -- Black hole physics -- X-rays: binaries -- methods: data analysis -- stars: individual: 4U 1543-47
\end{keywords}

\section{Introduction}

Transient black hole binary (BHB) systems are known to exhibit recurrent outbursts that can last for several months to years before going into period of quiescence. Throughout the outburst, most of these systems are found to evolve into different spectral states with contribution from mainly two emission components; soft thermal emission peaking at few keVs and hard powerlaw-like tail \citep{remillard2006x}. Based on their differing contributions, and timing features, several spectral states have been characterized for BHBs (for detailed classification see for e.g. \cite{states_belloni2011black}). At the start of an outburst, the system is found in Low/Hard State (LHS) in which the spectra can be described by a powerlaw component with slope $<$~2, which is expected to arise from Compton up-scattering of soft photons by a hot cloud of electrons. In LHS, the power spectrum shows strong variability with presence of band-limited noise \citep{husain2022detection} and sharp QPOs. As the outburst progresses, source can transition into what is defined as the High/Soft State (HSS), which in contrast to LHS, can be described by dominating thermal emission from the optically thick and geometrically thin accretion disk extending up to the Innermost Stable Circular Orbit (ISCO). This can also be accompanied by a weak high energy tail \citep{weak_tail_ss_capitanio2009}. In HSS, the source exhibits little variability with absence of sharp features in the power spectrum. The HSS is often noted to have a constant inner accretion disk \citep{ebisawa1993spectral} but an analysis of 4U 1957+115 by \cite{mudambi2022spectral} indicated an evolution of the inner edge in the HSS, hinting that the accretion disk extending to ISCO in this state may not be sacrosanct.  Observations conducted in HSS have also been used to detect disk winds \citep{wind_in_ss_miller, wind_in_ss_ponti}. Between LHS and HSS, two more intermediate states have been identified; Hard Intermediate and Soft Intermediate states (HIMS and SIMS) mainly based on characteristics of variability and the relative contributions of the disk and comptonised emission.

\bh\ is one such low-mass transient system which has been found to evolve through different states during an outburst, it has a secondary star of early spectral type, as reported in \cite{companion_star}. Many X-ray outbursts have been recorded from the transient since its first activity in 1971 \citep{first_detection_1972} each separated by 10--12~years (1983; \cite{kitamoto1984transient}, 1992; \cite{harmon1992}, 2002; \cite{park2004}) with the exception of latest outburst of 2021 which occurred after a time span of $\sim$20~years since its last recorded outburst in 2002. Based on dynamical measurements, \cite{orosz2002revised,orosz2003} estimated the mass of the black hole in \bh\ at $M_{BH}$=9.4$\pm$2.0~$M_{\odot}$. They also found it to be a low-inclination source with \textit{i}=20.7 $\pm$ $1.5^{\degree}$. Multiple studies using RXTE observations of the 2002 outburst have reported a broad range of spin values, for e.g. \cite{shafee2005spin} found a moderate spin range of 0.75--0.85 through continuum fitting whereas \cite{morningstar2014spin} reported a lower spin value of $\sim$0.43 and disk inclination angle of $32^{\degree}$. Although, \cite{morningstar2014spin} note that keeping the disk inclination fixed to $21^{\degree}$ the values found for spin are consistent with those found in \cite{shafee2005spin}. Latest spin study on 4U 1543-47 was performed by \cite{nustar_paper2022} 
 (hereafter \citetalias{nustar_paper2022}), where the authors through detailed reflection modeling determined that the black hole has a high spin of $0.98^{+0.01}_{-0.02}$. 

\textit{MAXI/GSC} observed the source on 2021 June $11^{\rm th}$ \citep{atel_14701} as it went into activity. This outburst is the brightest one detected for BHB systems in X-ray with its flux evolving to $\sim$5.4~Crab in 2.0--10.0~keV \citep[As seen with \textit{NICER},][]{atel_14725}. The source flux continued to rapidly increase soon after detection and reached a value close to its Eddington luminosity \citep{atel_14708}. The source transitioned to HSS as soon as it was detected and stayed in it until on 2022 January $3^{\rm rd}$ it was reported to have transitioned back to LHS. It was suggested by the hardening of \textit{MAXI/GSC} flux and radio brightening, which hints towards the  presence of a compact jet \citep{atel_15157}. Following the outburst, few small scale reflaring activities were also detected for the system on 2022 January $15^{th}$ \citep{atel_15253} and 2022 October $21^{\rm st}$ \citep{atel_15715}. 

\citetalias{nustar_paper2022} performed spectroscopy of \bh\ utilizing four \textit{NuSTAR} observations of its 2021 outburst, all extracted during HSS. They find strong presence of relativistic reflection (especially a strong dip in 7.0--11.0~keV) which could not be properly modelled even with six different flavours of \textit{Relxill} \citep{xillvergarcia2010x, relxill2014} family. The authors model the absorption by employing a warm absorber with the \textit{Relxill} component describing both comptonised and reflected emission. Similar to \citetalias{nustar_paper2022}, \cite{prabhakar2023} detected strong absorption via broadband spectral study using \textit{NuSTAR}, \textit{NICER} and \textit{AstroSat} data of the same outburst. To model the absorption, the authors implemented a gaussian absorption profile and  investigated multiple explanations (e.g. stellar winds, neutral cloud absorption, and disk winds) to explain its origin. The non-dependence of the absorption feature with the binary orbit phase rejected the stellar wind approach, while a  neutral partial absorber could not explain the spectra adequately and thus the authors suggest that the absorption was caused likely due to a disk wind.

In this study, we investigate the spectral properties of \bh\ by analysing a series of \textit{AstroSat} observations  during the decay phase of its 2021 outburst. Our analysis is centered on investigating the spectral evolution of the source while considering various factors responsible for this evolution. In Section 2, we discuss data reduction techniques for SXT and LAXPC data and in Section 3, we describe the broadband spectral analysis with various model configurations. Lastly, in Section 4, we discuss the results, along with potential interpretations of the observed spectral changes.

\section{Observations and Data reduction}
\label{sec:Data}

\begin{figure*}
    \centerline{\includegraphics[trim=0.5cm 0cm 1.5cm 0cm, clip=true, width=0.52\linewidth, height=5.5cm]{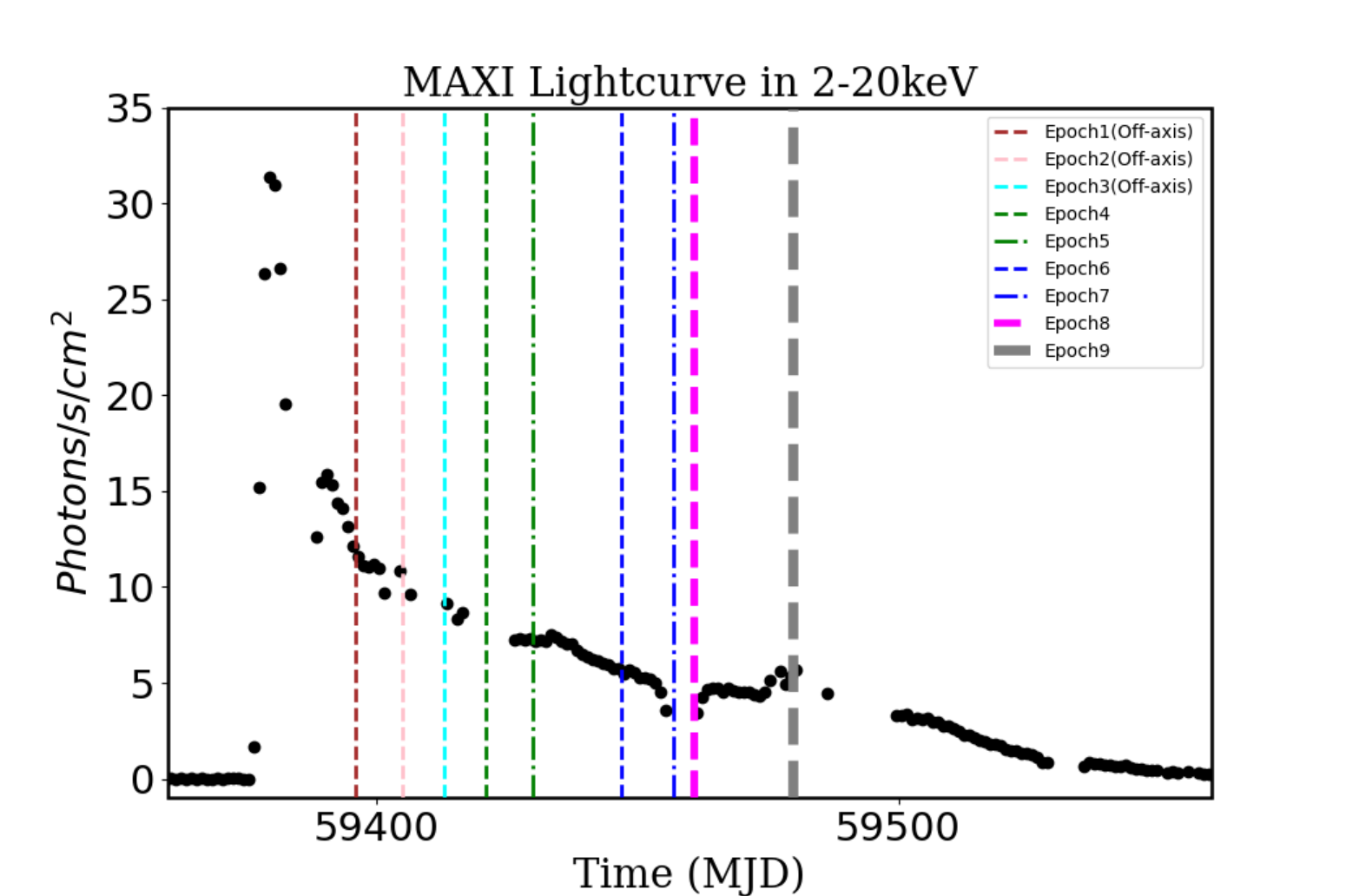}
    \includegraphics[trim=0.5cm 0cm 2cm 0cm, clip=true,width=0.53\linewidth, height=5.5cm]{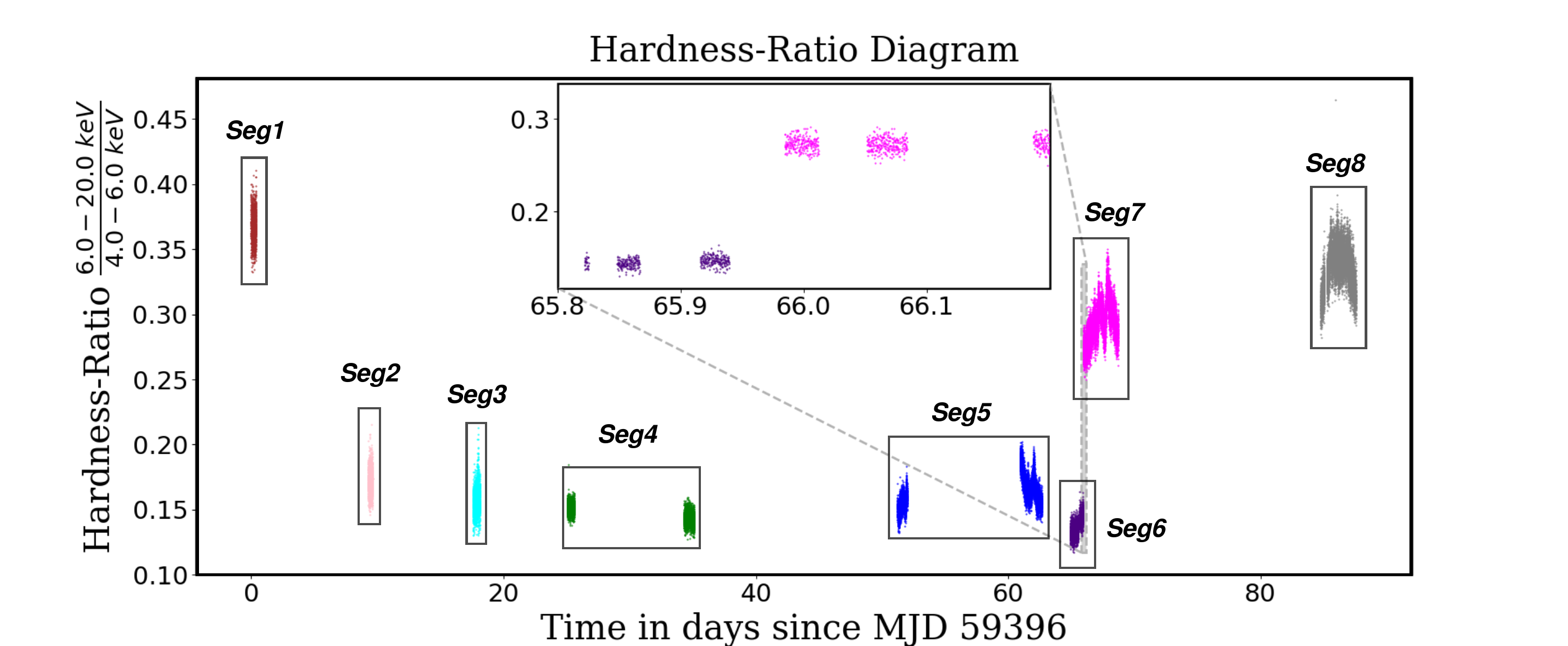}}
    \caption{In left panel; \textit{MAXI} lightcurve in 2.0--20.0~keV of \bh\  with vertical lines indicating the times of simultaneous \textit{AstroSat} observations. The first three observations (Epoch1--3) are off-axis  while all other observations (Epoch4--9) are on-axis. In right panel; we show Hardness-Ratio for 6.0--20.0/4.0--6.0~keV and segmentation of data. The data is divided into total of eight segments, each outlined with a box. The off-axis segments are labelled Seg1, Seg2, and Seg3 and are denoted by the colors brown, pink, and cyan, respectively. Epochs 3 and 4 are combined into Seg4, which is colored green, while Epochs 5 and 6 are grouped into Seg5, represented by the color blue. Epoch 8 is divided into two segments, Seg6 and Seg7, which are colored indigo and magenta, respectively. Epoch 9 is labelled Seg8 and is represented by the color grey. MJD 59396 denotes the start of Epoch1. Furthermore, we also show zoomed in plot for Epoch 8 which illustrates the period in which hardness-ratio of the source increased.}
    \label{fig:maxi}
\end{figure*}

On 2021 June $11^{\rm th}$, \textit{MAXI/GSC} nova alert system reported on a bright X-ray enhancement from the source indicating an onset of an outburst \citep{atel_14701}. Following it, we proposed  a series of  Target of Opportunity (ToO) observations of \bh\ with \textit{AstroSat} covering a period of roughly three months from July,~$1^{\rm st}$ to September,~$26^{\rm th}$ 2021. \textit{AstroSat} provides an opportunity to study such X-ray sources in broad band energy range of 0.3--80.0~keV with its two co-aligned instruments; Soft X-ray Telescope (SXT, 0.3--8.0~keV) and Large Area X-ray Proportional Counter (LAXPC, 4.0--80.0~keV). Details of these multi-epoch observations are given in Table~\ref{tab:log}. Even after a month of its outburst's detection, the source flux remained exceedingly high at $>2$~Crab, and thereby unsafe for on-axis observation with LAXPC (private communication with the POC) and therefore, the first three observations (Epoch 1 to 3) were taken by considering an off-set angle of 40$\arcmin$ with LAXPC. Later on, as the source flux decayed, the count rate became acceptable for on-axis observations, LAXPC and SXT both observed the source on six more occasions (Epoch 4 to 9). SXT data was not used for Epoch 1-3, given the source was largely at offset with SXT with it being out of the detector's field of view. We show all Epochs in left panel of Figure~\ref{fig:maxi} where we have plotted the \textit{MAXI} lightcurve (2.0--20.0~keV) of the whole outburst for \bh\ marking the \textit{AstroSat} observations as vertical lines.

\begin{table*}
\centering
\caption{Observation log for \textit{AstroSat} observations}    \renewcommand{\arraystretch}{0.8}{\begin{tabular}{p{2cm}|p{4cm}|p{2cm}|p{2cm}|p{2cm}}
\hline
    & Observation ID  & Start Time (UTC)  & End Time (UTC)  &  Exposure LAXPC/SXT (ksec) \\
\hline
     Epoch 1 &  $T04\_018T01\_9000004494$  & 2021-07-01  & 2021-07-01  &  16.4 (off-axis)\\
     & & 02:36:54  & 11:52:19 & \\
     Epoch 2 &  $T04\_021T01\_9000004526$  & 2021-07-10  & 2021-07-10  & 18.1 (off-axis)\\
     & & 08:07:04 & 18:15:54 & \\
     Epoch 3 &  $T04\_027T01\_9000004552$  & 2021-07-18  & 2021-07-19   & 17.7 (off-axis)\\
     & & 15:04:29 & 05:19:55 & \\
     Epoch 4 &  $T04\_030T01\_9000004588$  & 2021-07-26  & 2021-07-26   & 18.9/11.9\\
     & & 04:01:10 & 16:25:37 & \\
     Epoch 5 &  $T04\_035T01\_9000004622$  & 2021-08-04  & 2021-08-05   & 28.5/10.3 \\
     & & 10:24:02 & 04:58:24 & \\
     Epoch 6 &  $T04\_042T01\_9000004650$  & 2021-08-21 	& 2021-08-22  &  29.2/14.1\\
     & & 08:33:35 & 02:46:42 & \\
     Epoch 7 &  $T04\_046T01\_9000004680$ & 2021-08-31  &	2021-09-01  & 64.3/25.3\\
     & & 00:26:40 & 09:09:49  & \\
     Epoch 8 &  $T04\_051T01\_9000004686$  & 2021-09-04	& 2021-09-07   & 120.7/51.2 \\
     & &  00:15:34 & 21:41:56 & \\
     Epoch 9 &  $T04\_059T01\_9000004704$  & 2021-09-23 	& 2021-09-26  & 96.5/26.3\\
     & & 20:03:01 & 18:30:45  & \\
\hline \hline
\end{tabular}}
\label{tab:log}
\end{table*}

\begin{figure}
    \centerline{
    \includegraphics[trim=1.8cm 0cm 0cm 0cm, clip=true, width=9cm, height=6.5cm]{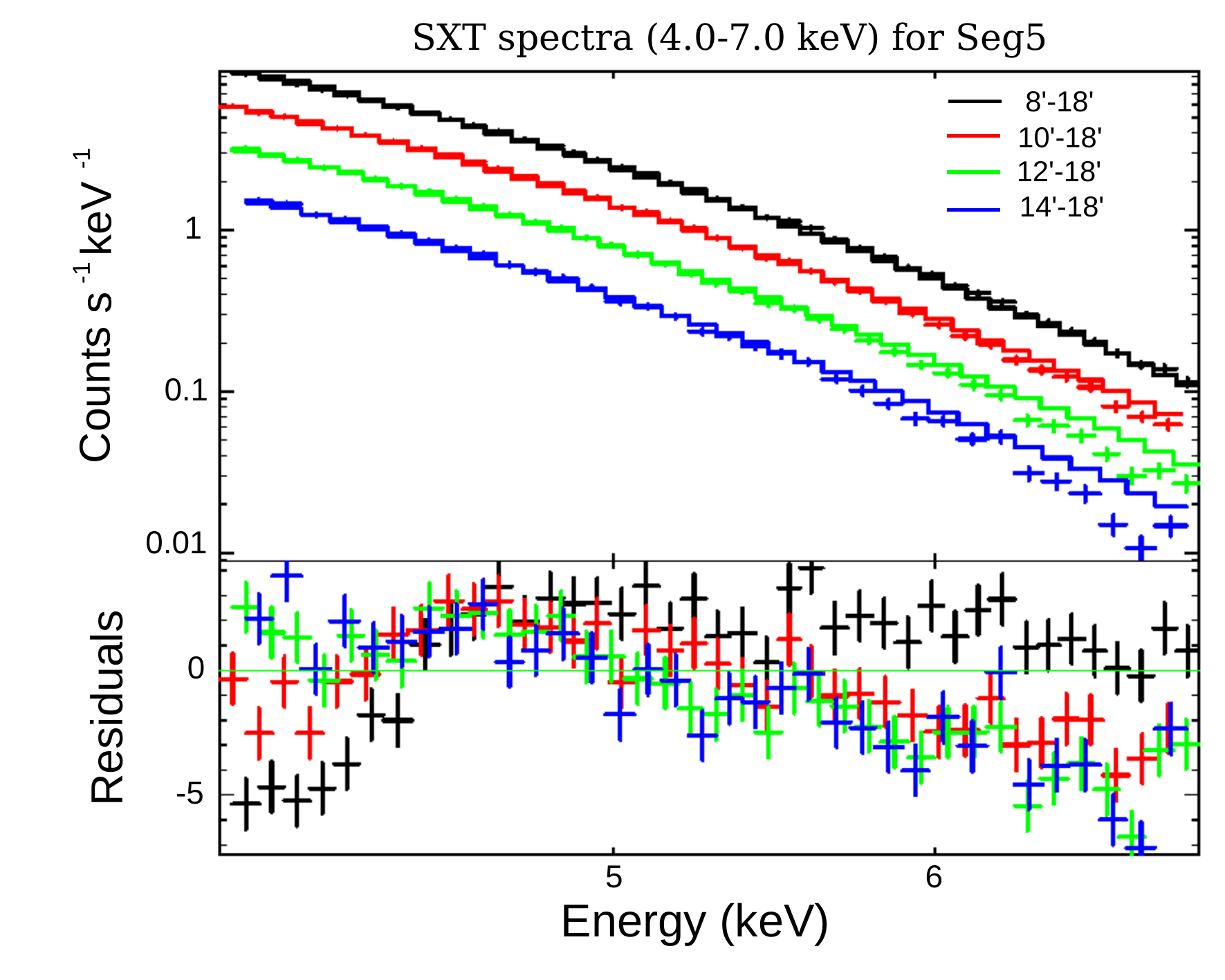}}
    \caption{Spectral variation shown for different inner radii of the annular region with outer region fixed to 18 arcmin to check for pile-up effects in SXT data. The joint fitting is performed with an absorbed powerlaw with photon index of all spectra tied. The residuals are shown in the lower panel.}
    \label{fig:pileup}
\end{figure}

\subsection{LAXPC}
LAXPC \citep{laxpc_agrawal2017,laxpc_antia2017calibration} has three identical proportional counters (\textit{LAXPC10}, \textit{LAXPC20} and \textit{LAXPC30}). It has a good timing resolution of 10~$\mu$s. Currently, \textit{LAXPC20} remains the only counter working nominally as \textit{LAXPC30} suffered from gas leakage and abnormal gain change was reported for \textit{LAXPC10} early in 2018, therefore in this work we utilize data from only \textit{LAXPC20}.  The data reduction from Level-1 to Level-2 was carried out with \textit{LAXPC Software\footnote{LAXPC software was obtained from http://astrosat-ssc.iucaa.in/laxpcData}}. Good time intervals \textit{`GTI'} for the data were created using `\textit{laxpc$\_$make$\_$stdgti}' which excludes duration of the passage over South Atlantic Anomaly (SAA) and Earth occultation of the source. GTI was further corrected by removing the slew operation time of the satellite. To study the behaviour of source, we also make use of the Hardness-Ratio (HR) diagram with LAXPC data, which is defined here as the ratio of flux in the hard band (6.0--20.0~keV) to soft band (4.0--6.0~keV). We depict the variation of the HR as a function of time in right panel of Figure~\ref{fig:maxi}. 

Using the evolution of the HR we divide our observations into eight segments. Each segment is numbered and enclosed in a box as shown in right panel of Figure~\ref{fig:maxi}. We have segmented the off-axis Epochs into distinct segments with each segment represented by a different color: Seg1 in brown, Seg2 in pink, and Seg3 in cyan. We have combined Epochs 4 and 5, which exhibited similar hardness, into Seg4 represented in green. Similarly, we have merged Epochs 6 and 7 into Seg5 represented in blue. However, Epoch 8 displayed variations in hardness during the observation, therefore we have split it into two segments: Seg6 and Seg7 represented in indigo and magenta, respectively. Finally, Epoch 9 is represented by a single segment, Seg8, depicted in grey. For each segment we extract the LAXPC spectra with subroutine `\textit{laxpc$\_$make$\_$spectra}' and Background spectra with `\textit{laxpc$\_$make$\_$backspectra}', with corresponding Response Matrix File (RMF) being generated along with the spectra. For the off-axis segments, we considered response files created with an offset angle of 40$\arcmin$ provided separately by LAXPC POC team \footnote{https://www.tifr.res.in/~astrosat$\_$laxpc/LaxpcSoft.html}.\\

\subsection{SXT}

SXT \citep{sxt2016orbit,sxt2017soft} is a focusing X-ray telescope with a focal plane camera to detect X-rays in 0.3--8.0~keV. All SXT on-axis observations of \bh\ were conducted in Photon Counting (PC) mode. We use the Level2 SXT data which was already processed through SXT pipeline\footnote{sxtpipeline1.4b (Release Date: 2019-01-04)}. The cleaned events of each orbit were merged into a single event file for each segment using Julia-based merger tool `\textit{SXTMerger tool}', this removes any overlapping events in successive orbits. 

We then extracted the SXT science products in XSelect (V2.4m). SXT data suffered from large pile-up since the SXT count rate was much higher than the threshold value of pile-up ($<40~Counts/s$) for PC mode. To evaluate the presence of spectral distortion due to pile-up, we examine an annular region that has a fixed outer radius of 18 arcmin, while we vary the inner radius up to a maximum of 14 arcmin. The SXT spectra extracted for different inner radii are modelled simultaneously by an absorbed powerlaw with their photon index tied and allowing normalization to be free. As shown in Figure~\ref{fig:pileup} for Seg5, we find the distortion was largely removed for annular region of 12--18 arcmin (given the similarity of residuals seen for 12--18 (green) and 14--18 armin (blue) regions), which we utilize to extract the final products. The final spectra was rebinned using the command line tool \textit{ftgrouppha} using optimal binning. Also, the standard ancillary response file (sxt$\_$pc$\_$excl00$\_$v04$\_$20190608$\_$mod$\_$16oct21.arf) was corrected for the source region selection and vignetting effect which is due to an off-set between the source position and the optical-axis of SXT, this was accomplished by \textit{`SXTARFModule'}. Finally, the rebinned SXT spectrum was combined with the corrected ARF file, response file `\textit{sxt$\_$pc$\_$mat$\_$g0to12.rmf}' and deep background spectrum `\textit{SkyBkg$\_$comb$\_$EL3p5$\_$Cl$\_$Rd16p0$\_$v01.pha}' provided by the SXT team\footnote{http://astrosat-ssc.iucaa.in/sxtData}.

\section{Spectral Analysis}
\label{sec:Analysis}

Here, we present fitting results for both off- and on-axis observations separately. To begin with, we analyse the spectra taken for on-axis observations (Seg4 to Seg8) given the broad energy coverage of spectra with both LAXPC and SXT to determine best-fit models. We use these best-fit models to investigate spectral characteristics of off-axis observations (Seg1 to Seg3) in subsection 3.2.

\subsection{On-axis observations}

    \begin{figure*}
   \hspace{-0.8cm}
    \includegraphics[trim=0 0 0 0cm,clip, width=10.5cm, height=12cm]{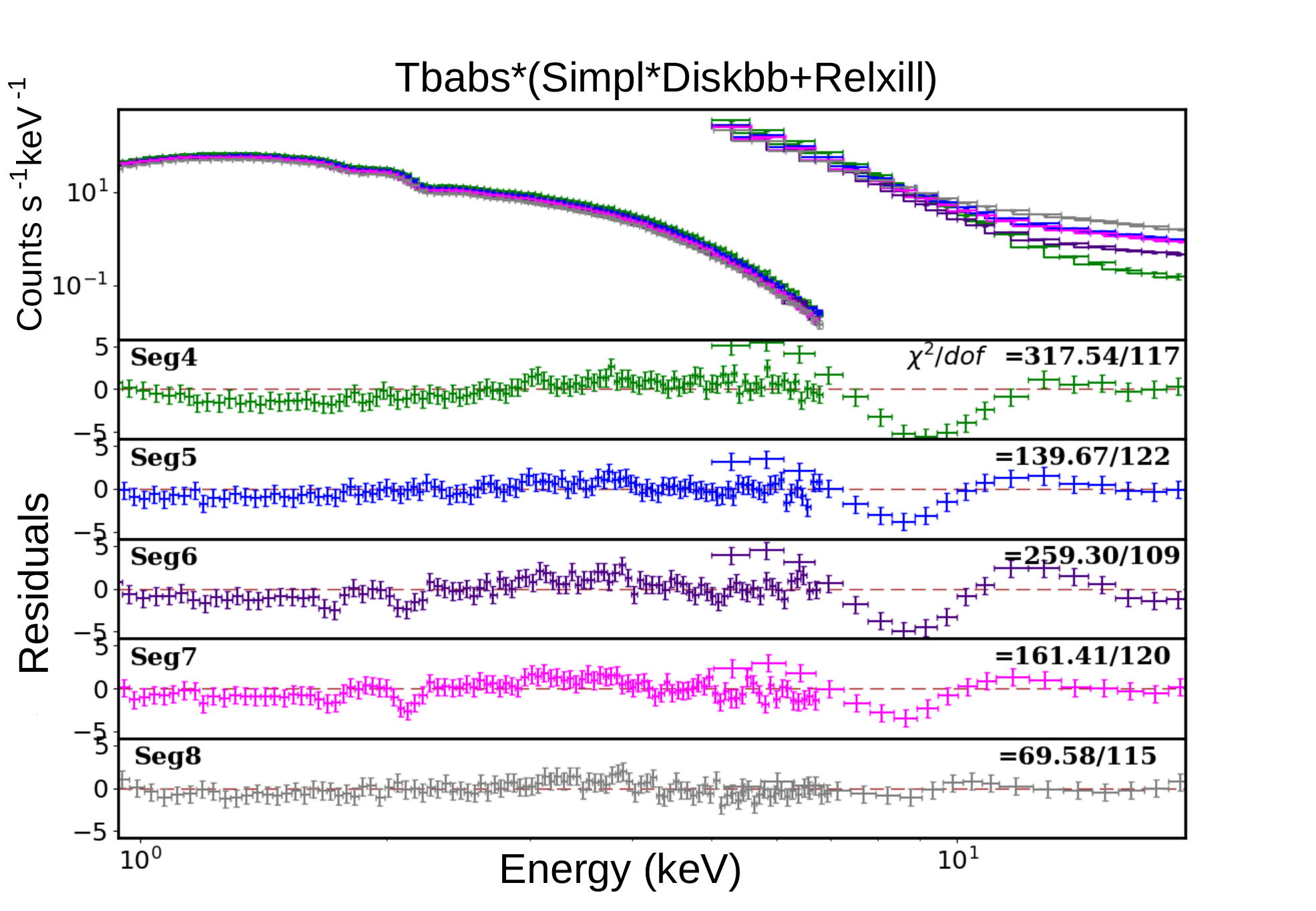}\hfill
    \hspace{-4.cm}
    \includegraphics[trim=1.1cm 0 0 0cm,clip, width=9.9cm, height=12cm]{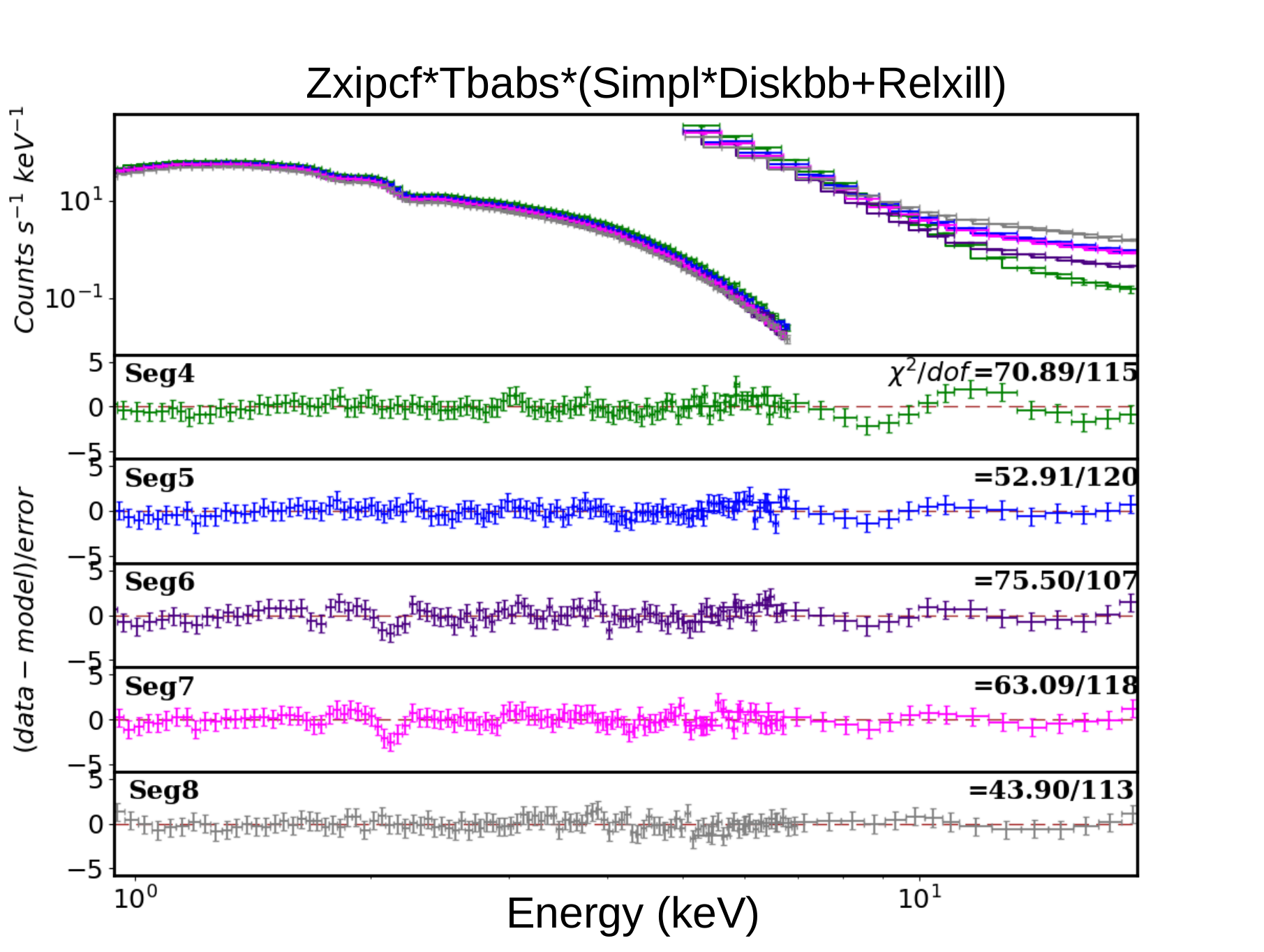}\hfill
       \hspace{-4.cm}
   \caption{Data and residuals for model \textit{TBabs(Simpl$\otimes$Diskbb+Relxill)} are shown in Left Panel, and after including  \textit{Zxipcf} i.e. \textit{Zxipcf*TBabs*(Simpl$\otimes$Diskbb+Relxill)} are shown in Right panel. Each subpanel also has respective value of $\chi^2/dof$. It follows the color scheme of Figure~1; right panel.}
   \label{fig:delchi_dbb_kerrd}
\end{figure*}

\begin{table*}
\renewcommand{\arraystretch}{0.3}
\tabcolsep=0.2cm 
\caption{Best fit results for Model1, Model2 and Model3 for all segments}
\begin{threeparttable}
\centerline{\begin{tabular}{l l l l l l l l l l l}

\hline

\multicolumn{11}{c}{\textit{\Large{Const$\times$Zxipcf$\times$TBabs$\times$(Relxill+Simpl$\otimes$Diskbb)}}} \\

\hline

Model & Parameter & Unit & Seg1 & Seg2 & Seg3 & Seg4 & Seg5 & Seg6 & Seg7 & Seg8 \\

\hline 

& \textit{Const} & & & & & $1.52^{+0.05}_{-0.05}$ & $1.56^{+0.05}_{-0.05}$ & $1.31^{+0.05}_{-0.05}$ & $1.37^{+0.05}_{-0.05}$ & $1.35^{+0.05}_{-0.05}$ \\ [1ex]

\textit{Zxipcf} & $n_{H}$ & $10^{22}$ cm$^{-2}$ & 13.50 & 13.50 & 13.50 & $13.50^{+1.65}_{-1.77}$ & $11.93^{+2.28}_{-2.53}$ & $11.67^{+1.77}_{-1.85}$ & $9.45^{+2.41}_{-3.09}$ & $8.27^{+1.66}_{-4.37}$ \\ [1ex]

& $log(\xi)$ & -- & -- & -- & -- & -- & $2$ (fixed) & -- & -- & -- \\ [1ex]

& $cf$ & & 0.44 & 0.44 & 0.44 & $0.44^{+0.03}_{-0.04}$ & $0.36^{+0.06}_{-0.07}$ & $0.45^{+0.05}_{-0.05}$ & $0.34^{+0.07}_{-0.06}$ & $0.21^{+0.07}_{-0.04}$ \\ [1ex]

\textit{TBabs} & $n_{H}$ & $10^{22}$ cm$^{-2}$ & 0.50 & 0.50 & 0.50 & $0.50^{+0.03}_{-0.02}$ & $0.51^{+0.04}_{-0.04}$ & $0.45^{+0.02}_{-0.02}$ &  $0.46^{+0.03}_{-0.03}$ & $0.53^{+0.02}_{-0.03}$ \\ [1ex]

\textit{Relxill} & $N_{relx}$ &  & $1.242^{+0.146}_{-1.133}$  & $0.722^{+0.242}_{-0.717}$ & $0.010^{+0.040}$  & $0.156^{+0.040}_{-0.029}$ &  $0.045^{+0.040}_{-0.029}$ & $0.005^{+0.009}$ & $0.034^{+0.033}_{-0.021}$ & $0.086^{+0.053}_{-0.046}$ \\ [1ex]

\textit{Simpl} & $\Gamma$ &  & $3.43_{-2.33}$ & $3.23_{-0.60}$ & $3.58_{-0.63}$ & $3.40^{+0.90}_{-0.75}$ & $2.77^{+0.32}_{-0.19}$ & $2.50^{+0.41}_{-0.31}$ & $2.73^{+0.26}_{-0.20}$ & $2.78^{+0.16}_{-0.18}$  \\ [1ex]
  
& $f_{sc}$ & & -- & -- & -- & -- & $0.001$ (fixed) & -- & -- & -- \\ [1ex]
  
\textit{Diskbb} & $kT_{in}$ & keV & $0.93^{+0.01}_{-0.01}$ & $0.85^{+0.01}_{-0.01}$ & $0.81^{+0.01}_{-0.01}$ & $0.84^{+0.01}_{-0.01}$ & $0.84^{+0.01}_{-0.02}$ & $0.78^{+0.01}_{-0.01}$ & $0.82^{+0.01}_{-0.02}$ & $0.82^{+0.02}_{-0.01}$ \\ [1ex]

& $N_{Disk}$ & & $15538^{+1781}_{-1642}$ & $20249^{+1934}_{-2391}$ & $21820^{+3037}_{-2801}$ & $9583^{+1127}_{-574}$ & $7667^{+917}_{-809}$ & $12363^{+1261}_{-1534}$ & $8396^{+828}_{-952}$ & $6425^{+963}_{-800}$ \\ [1ex]

& $R_{in}$\tnote{a} & $R_g$ & $20.2\pm1.1$ & $23.0\pm1.2$  & $23.9\pm1.6$ & $15.8\pm0.7$ & $14.2\pm0.8$ & $17.8\pm1.0$ & $14.8\pm0.8$ & $13.0\pm0.9$ \\ [1ex]

& $R^{'}_{in}$\tnote{b} & $R_g$ & $7.0\pm0.4$ & $8.0\pm0.4$ & $8.3\pm0.6$ & $5.5\pm0.2$ & $4.9\pm0.3$ & $6.2\pm0.4$ & $5.1\pm0.3$ & $4.5\pm0.3$ \\ [1ex]

\hline

& $\chi^2/dof$ & & 19.67/14 & 11.45/14 & 12.5/14 & 70.89/115 & 52.91/120 & 75.5/107 & 63.09/118 & 44.80/113 \\
& & & =1.41 & =0.82 & =0.89 & =0.62 & =0.44 & =0.71 & =0.53 & =0.40 \\ [1ex]

&\textit{$F_{abs}$\tnote{c}} & $10^{-8}$ erg $cm^{-2} s^{-1}$ & $10.15$ & $9.54$ & $8.01$ & $4.23$ & $3.59$ & $3.85$ & $3.71$ & $3.29$\\ [1ex]

&\textit{$L/L_{Edd}$} &  & $0.578$ & $0.543$ & $0.456$ & $0.241$ & $0.204$ & $0.219$ & $0.211$ & $0.204$\\ [1ex]

&\textit{$F_{ubs}$\tnote{d}} & $10^{-8}$ erg $cm^{-2} s^{-1}$ & $20.88$ & $18.89$ & $16.07$ & $8.43$ & $6.55$ & $7.85$ & $6.47$ & $5.26$\\ [1ex]

\hline 

\multicolumn{11}{c}{\textit{\Large{Const$\times$Zxipcf$\times$TBabs$\times$(Relxill+Simpl$\otimes$Kerrd)}}} \\

\hline


& \textit{Const} & & & & & $1.51^{+0.06}_{-0.05}$ & $1.58^{+0.04}_{-0.04}$ & $1.29^{+0.05}_{-0.04}$ & $1.39^{+0.02}_{-0.05}$ & $1.37^{+0.05}_{-0.04}$ \\ [1ex]

\textit{zxipcf} & $n_{H}$ &  $10^{22}$ cm$^{-2}$ & 12.03 & 12.03 & 12.03 & $12.03^{+0.92}_{-1.19}$ & $11.11^{+1.22}_{-1.00}$ & $11.84^{+1.40}_{-1.31}$ & $8.48^{+2.39}_{-1.62}$ & $8.29^{+0.72}_{-2.82}$ \\ [1ex]

& $log(\xi)$ & -- & -- & -- & -- & -- & $2$ (fixed) & -- & -- & -- \\ [1ex]

& $cf$ & & 0.53 & 0.53 & 0.53 & $0.53^{+0.02}_{-0.05}$ & $0.45^{+0.05}_{-0.02}$ & $0.57^{+0.02}_{-0.02}$ & $0.41^{+0.08}_{-0.02}$ & $0.34^{+0.02}_{-0.04}$ \\ [1ex]

\textit{TBabs} & $n_{H}$ & $10^{22}$ cm$^{-2}$ & 0.54 & 0.54 & 0.54 & $0.54^{+0.02}_{-0.02}$ & $0.60^{+0.03}_{-0.03}$ & $0.55^{+0.04}_{-0.03}$ & $0.54^{+0.03}_{-0.03}$ & $0.61^{+0.02}_{-0.04}$ \\ [1ex]

\textit{Relxill} & $N_{relx}$ & & $0.356_{-0.316}$ & $0.216^{+0.358}_{-0.210}$ & $0.081^{+0.132}$  & $0.112^{+0.023}$ & $0.101^{+0.098}_{-0.061}$ & $0.099^{+0.141}_{-0.076}$ & $0.063^{+0.054}_{-0.034}$ & $0.136^{+0.068}_{-0.061}$ \\ [1ex]

\textit{Simpl} & $\Gamma$ & & $3.22_{-0.44}$ & $3.20_{-0.64}$ & $3.06_{-1.79}$ & $3.39^{+0.63}_{-0.85}$ & $2.99^{+0.16}_{-0.16}$ & $3.18_{-0.56}$ & $2.92^{+0.22}_{-0.16}$ & $2.94^{+0.08}_{-0.15}$  \\ [1ex]
  
& $f_{sc}$ & & -- & -- & -- & -- & $0.001$ (fixed) & -- & -- & -- \\ [1ex]
  
\textit{Kerrd} & $\dot{M}$ & $10^{18} g/s$ & $15.51^{+0.61}_{-0.84}$ & $16.79^{+0.70}_{-1.37}$ & $15.05^{+0.63}_{-0.82}$ & $6.12^{+0.42}_{-0.21}$ & $4.43^{+1.20}_{-1.20}$ & $7.17^{+0.43}_{-0.30}$ & $4.18^{+1.38}_{-0.02}$ & $3.34^{+0.30}_{-0.30}$ \\ [1ex]

& $R_{in}$ & $R_g$ & $15.72^{+0.27}_{-0.40}$ & $18.86^{+0.50}_{-0.64}$ & $19.36^{+0.43}_{-0.34}$ & $12.84^{+0.41}_{-0.41}$ & $11.22^{+0.87}_{-0.29}$ & $15.72^{+0.27}_{-0.28}$ & $11.13^{+0.38}_{-0.20}$ & $10.50^{+0.56}_{-0.67}$ \\ [1ex]

\hline

& $\chi^2/dof$ & & 22.2/14 & 10.7/14 & 10.8/14 & 78.64/115 & 52.8/120 & 76.5/107 & 63.47/118 & 55.21/113 \\
& & & =1.59 & =0.76 & =0.77 & =0.68 & =0.44 & =0.72 & =0.54 & =0.49 \\[1ex]

&\textit{$F_{abs}$} & $10^{-8}$ erg $cm^{-2} s^{-1}$ & $11.10$ & $9.51$ & $7.64$ & $4.23$ & $3.59$ & $3.83$ & $3.73$ & $3.26$\\ [1ex]

&\textit{$F_{unabs}$} & $10^{-8}$ erg $cm^{-2} s^{-1}$ & $23.89$ & $21.62$ & $17.66$ & $9.63$ & $7.68$ & $9.73$ & $7.32$ & $6.28$\\ [1ex]

\hline

\multicolumn{11}{c}{\textit{\Large{Const$\times$Zxipcf$\times$TBabs$\times$(Relxill+Simpl$\otimes$Kerrbb)}}} \\

\hline


& \textit{Const} & & & & & $1.52^{+0.05}_{-0.05}$ & $1.56^{+0.05}_{-0.05}$ & $1.29^{+0.05}_{-0.05}$ & $1.37^{+0.05}_{-0.05}$ & $1.36^{+0.05}_{-0.05}$ \\ [1ex]

\textit{Zxipcf} & $n_{H}$ & $10^{22}$ cm$^{-2}$ & 13.15 & 13.15 & 13.15 & $13.15^{+1.31}_{-1.43}$ & $11.30^{+1.81}_{-1.98}$ & $11.67^{+1.50}_{-1.56}$ & $9.44^{+1.91}_{-2.58}$ & $8.03^{+1.17}_{-3.23}$ \\ [1ex]

& $log(\xi)$ & -- & -- & -- & -- & -- & $2$ (fixed) & -- & -- & -- \\ [1ex]

& $cf$ & & 0.52 & 0.52 & 0.52 & $0.52^{+0.03}_{-0.03}$ & $0.43^{+0.05}_{-0.06}$ & $0.54^{+0.04}_{-0.04}$ & $0.42^{+0.06}_{-0.06}$ & $0.29^{+0.07}_{-0.05}$ \\ [1ex]

\textit{TBabs} & $n_{H}$ & $10^{22}$ cm$^{-2}$ & 0.53 & 0.53 & 0.53 & $0.53^{+0.02}_{-0.02}$ & $0.55^{+0.03}_{-0.02}$ & $0.48^{+0.02}_{-0.02}$ &  $0.50^{+0.03}_{-0.03}$ & $0.56^{+0.03}_{-0.03}$ \\ [1ex]

\textit{Relxill} & $N_{relx}$ &  & $1.024^{+0.067}_{-0.067}$  & $0.690^{+0.055}_{-0.055}$ & $0.520^{+0.049}_{-0.049}$  & $0.125^{+0.033}$ & $0.076^{+0.078}_{-0.056}$ & $0.005^{+0.001}$ & $0.045^{+0.033}_{-0.028}$ & $0.109^{+0.102}_{-0.044}$ \\ [1ex]

\textit{Simpl} & $\Gamma$ &  & 3.41 & 3.41 & 3.41 & $3.41^{+0.84}_{-0.89}$ & $2.92^{+0.25}_{-0.32}$ & $2.53^{+0.63}_{-0.60}$  & $2.82^{+0.31}_{-0.22}$ & $2.87^{+0.14}_{-0.12}$  \\ [1ex]
  
& $f_{sc}$ & & -- & -- & -- & -- & $0.001$ (fixed) & -- & -- & -- \\ [1ex]
  
\textit{Kerrbb} & $a$ & & & & & -- & 0.43 (fixed) & -- & -- & -- \\ [1ex]

& $M_{dd}$ & & $18.25^{+1.25}_{-1.13}$ & $17.37^{+1.41}_{-1.27}$ & $15.87^{+1.45}_{-1.29}$ & $7.87^{+0.52}_{-0.32}$ & $5.93^{+0.07}_{-0.07}$ & $7.67^{+0.74}_{-0.33}$ & $5.95^{+0.18}_{-0.47}$ & $4.56^{+0.37}_{-0.41}$ \\ [1ex]

& $f_{col}$ & & $1.42^{+0.04}_{-0.04}$ & $1.32^{+0.04}_{-0.04}$ & $1.27^{+0.05}_{-0.05}$ & $1.59^{+0.05}_{-0.02}$ & $1.69^{+0.05}_{-0.05}$ & $1.49^{+0.06}_{-0.03}$ & $1.65^{+0.06}_{-0.02}$ & $1.77^{+0.07}_{-0.07}$ \\ [1ex]

\hline

& $\chi^2/dof$ & & 19.17/15 & 11.57/15 & 13.75/15 & 74.34/115 & 55.68/120 & 79.51/107 & 64.43/118 & 48.93/113 \\
& & & =1.28 & =0.77 & =0.92 & =0.65 & =0.46 & =0.74 & =0.55 & =0.43 \\[1ex]

&\textit{$F_{abs}$} & $10^{-8}$ erg $cm^{-2} s^{-1}$ & $10.84$ & $9.50$ & $8.12$ & $4.22$ & $3.63$ & $3.90$ & $3.75$ & $3.30$\\ [1ex]

&\textit{$F_{unabs}$} & $10^{-8}$ erg $cm^{-2} s^{-1}$ & $23.29$ & $21.29$ & $18.97$ & $9.42$ & $7.28$ & $8.91$ & $7.20$ & $5.80$\\ [1ex]

\hline


\end{tabular}}
\begin{tablenotes}
\item [a] $R_{in}$ is estimated from $N_{Disk}$ using the expression $f_{col}^2 \times D_{10} \times(N_{disk}/cos(i))^{1/2}$ \citep{diskbb}.
\item [b] $R^{'}_{in}$ is estimated with no spectral hardening.
\item [c] $F_{abs}$ is total absorbed flux in 0.7-20.0~keV calculated using `\textit{Cflux}'. 
\item [d] $F_{unabs}$ is unabsorbed flux (without contribution from \textit{Zxipcf} and \textit{Tbabs}) in 0.7-20.0~keV calculated using `\textit{Cflux}'.

\end{tablenotes}
\end{threeparttable}
\label{tab:diskbb}
\end{table*}

We begin by fitting the combined spectra of SXT and LAXPC in energy range 0.7--20.0~keV individually for Seg4 to Seg8, we ignored below 0.7~keV due to response uncertainties and above 20.0~keV due to background domination. In order to assess the joint spectra in XSpec(v12.12.0), we use \textit{TBabs} \citep{TBabs} to model the galactic absorption along the line of sight considering set of solar abundances by  \citet{aspl} and photoelectric cross-sections by  \citet{vern}. To model the thermal continuum we implement \textit{Diskbb} \citep{diskbb} which considers an optically thick accretion disk emitting multi-color blackbody radiation giving inner disk temperature ($kT_{in}$) and  normalization ($N_{Disk}$) as fit parameters. For high energy tail emission we use the convolution component \textit{Simpl} \citep{simpl} which considers an up-scattered comptonisation component by taking disk emission as the seed spectrum and returns power law index ($\Gamma$) and fractional scattering ($f_{sc}$, fraction of photons up-scattered) as fit parameters. A cross-normalization factor (a constant) was also added given the spectra are from two different detectors. Accordingly, we individually fit the spectrum for each segment with the model \textit{TBabs*(Simpl$\otimes$Diskbb)} and find the source to be in soft state with major contribution coming from the thermal emission with $F_{Disk}/F_{Total}$ $>$ 95$\%$. We find large residuals in LAXPC data above 5.0~keV, including a significant absorption between 7.0-11.0~keV. These residuals are most pronounced in Seg4 and least noticeable in Seg8.

Apart from thermal and comptonized spectra, irradiation of the disk by up-scattered photons can also contribute to the spectra. This can result into reflection features like the Fe emission, Compton hump and absorption edges due to high fluorescence yield of Fe, which can be broadened and skewed owing to relativistic and GR effects in the inner region. But we find there is no visible line-like residual in 6.0--7.0~keV in the spectra, as reported earlier for this source by \cite{park2004}, therefore addition of a \textit{gaussian} does not lead to a significant improvement in fit. Since the comptonised emission is weak, the corresponding reflection is also expected to be weak. But to have a better estimate of the inner accretion disk properties the effects at harder X-rays have to be accurately modelled.

To incorporate the effects of the weak reflection, we include \textit{Relxill} \citep{xillvergarcia2010x, relxill2014} with limited free parameters. We tie the reflection photon index to the comptonising component's index. From hereon, inclination, distance and mass of the black hole are set to 21$\degree$, 7.5~kpc and 9.4~$M_{\odot}$ respectively, unless stated otherwise. The ionization parameter, $log(\xi)$ to 3 and Fe abundance are set to solar values. In agreement with \citetalias{nustar_paper2022}, we fixed the spin parameter in \textit{Relxill} to 0.998, since fixing it to different values did not affect the fit. The reflection fraction was fixed at -1. We try with fixing $R_f$ to different values but it doesn't affect the fit notably. We allowed only the normalization ($N_{relx}$) to vary keeping all other parameters fixed. Due to inclusion of reflection component, the statistics of the data do not allow simultaneous constraint on both  $f_{sc}$  and $N_{relx}$ and in some cases force $f_{sc}$ to take a non-physical value (e.g. $10^{-20}$ for Seg4), hence we fix $f_{sc}$ to a fiducial low value of 0.001 for each segment. Even after addition of \textit{Relxill}, the large absorption dip could not be accounted for as can be seen in left panel of Figure~\ref{fig:delchi_dbb_kerrd},  where we have plotted data and residuals for Model \textit{Tbabs*(Simpl$\otimes$Diskbb+Relxill)}, indicating a need for an additional absorption of high energy photons. We also mention the reduced $\chi^2$ for each segment in the corresponding panel. 

\begin{figure*}
    \hspace{-5cm}
    \includegraphics[width=7.5cm, height=5cm]{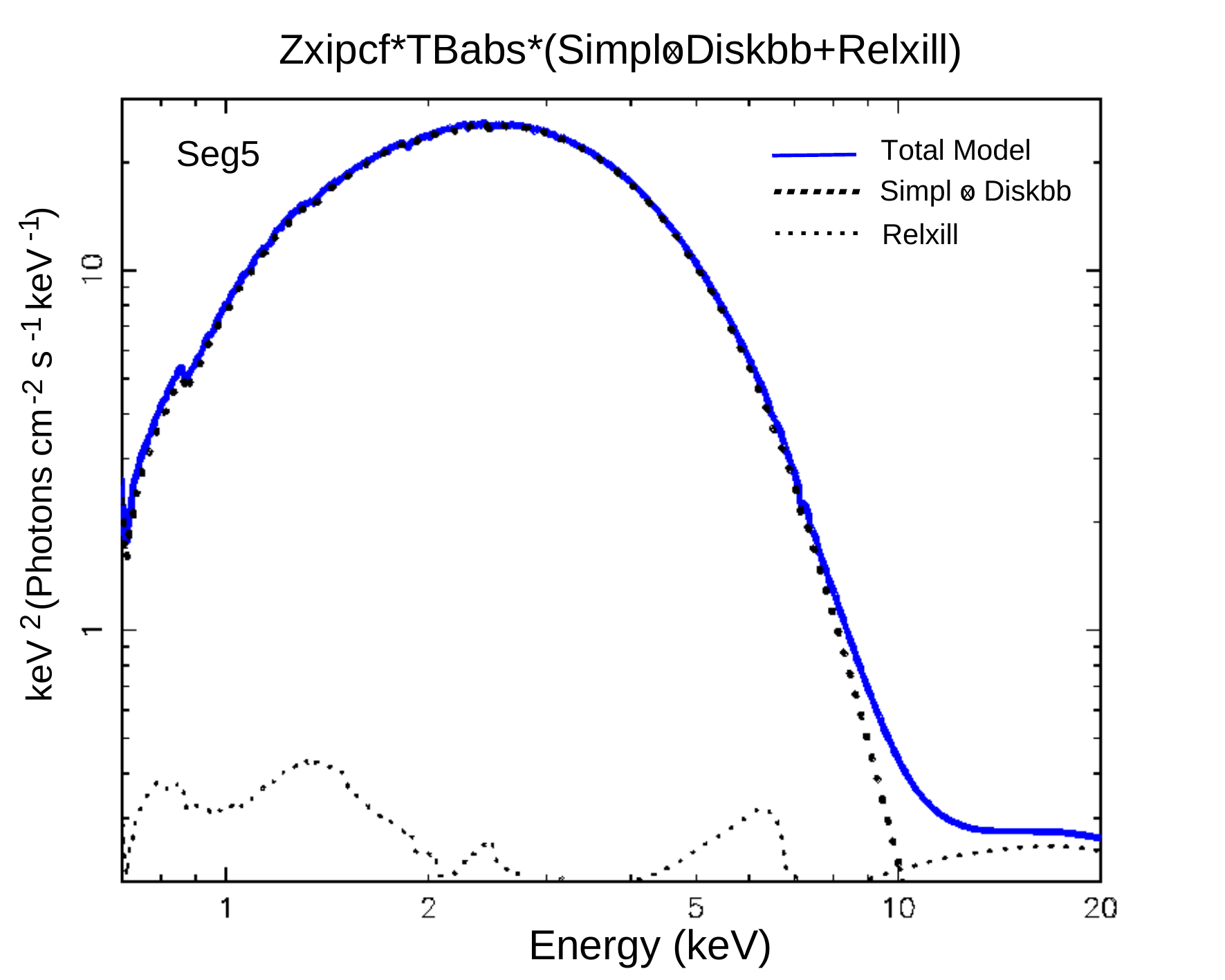} 
    \hspace{-0.5cm}
    \includegraphics[trim=1.5cm 0cm 0cm 0cm, clip=true, width=6.5cm, height=5cm]{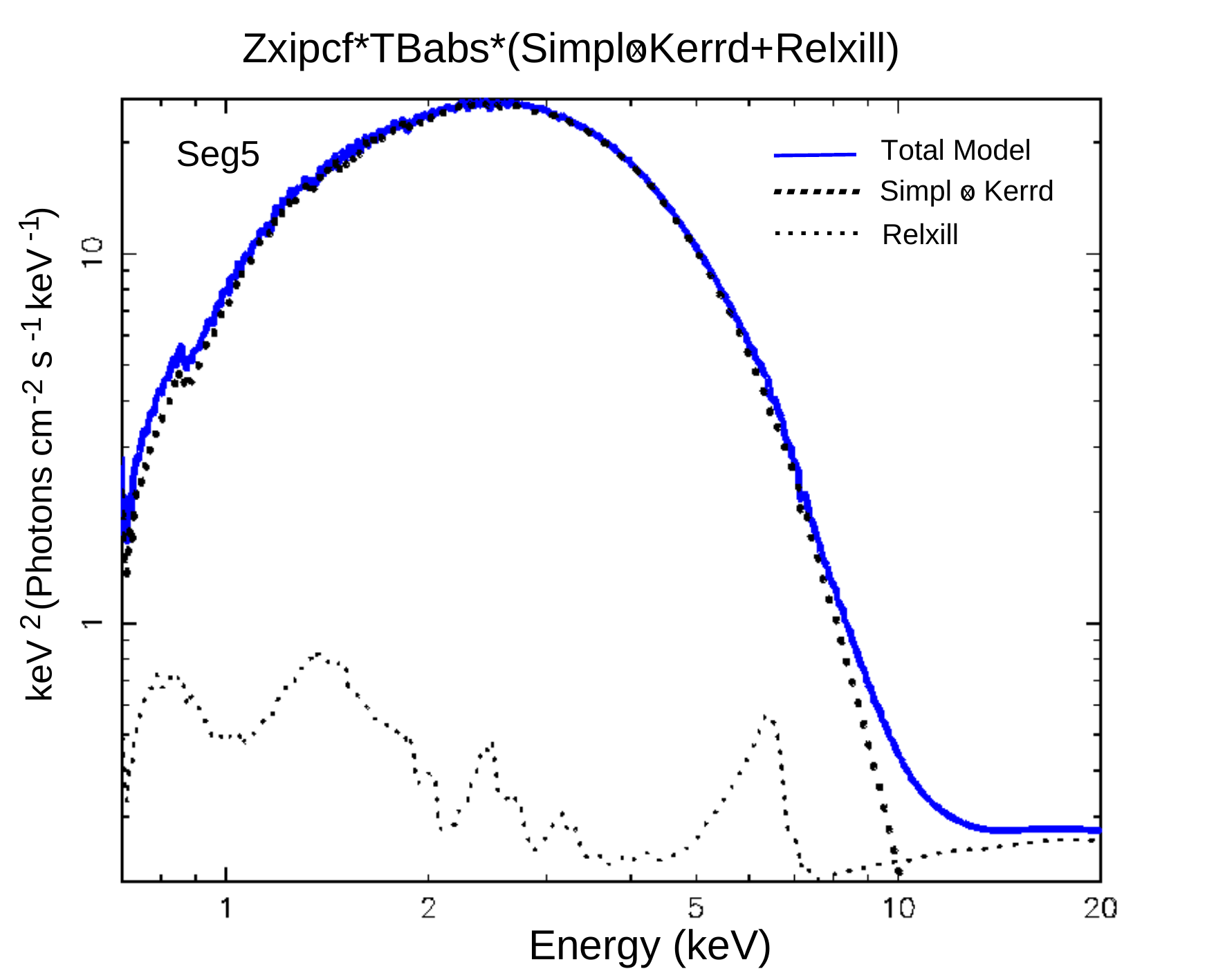}
     \hspace{-0.5cm}
    \includegraphics[trim=1.5cm 0cm 0cm 0cm, clip=true, width=6.5cm, height=5cm]{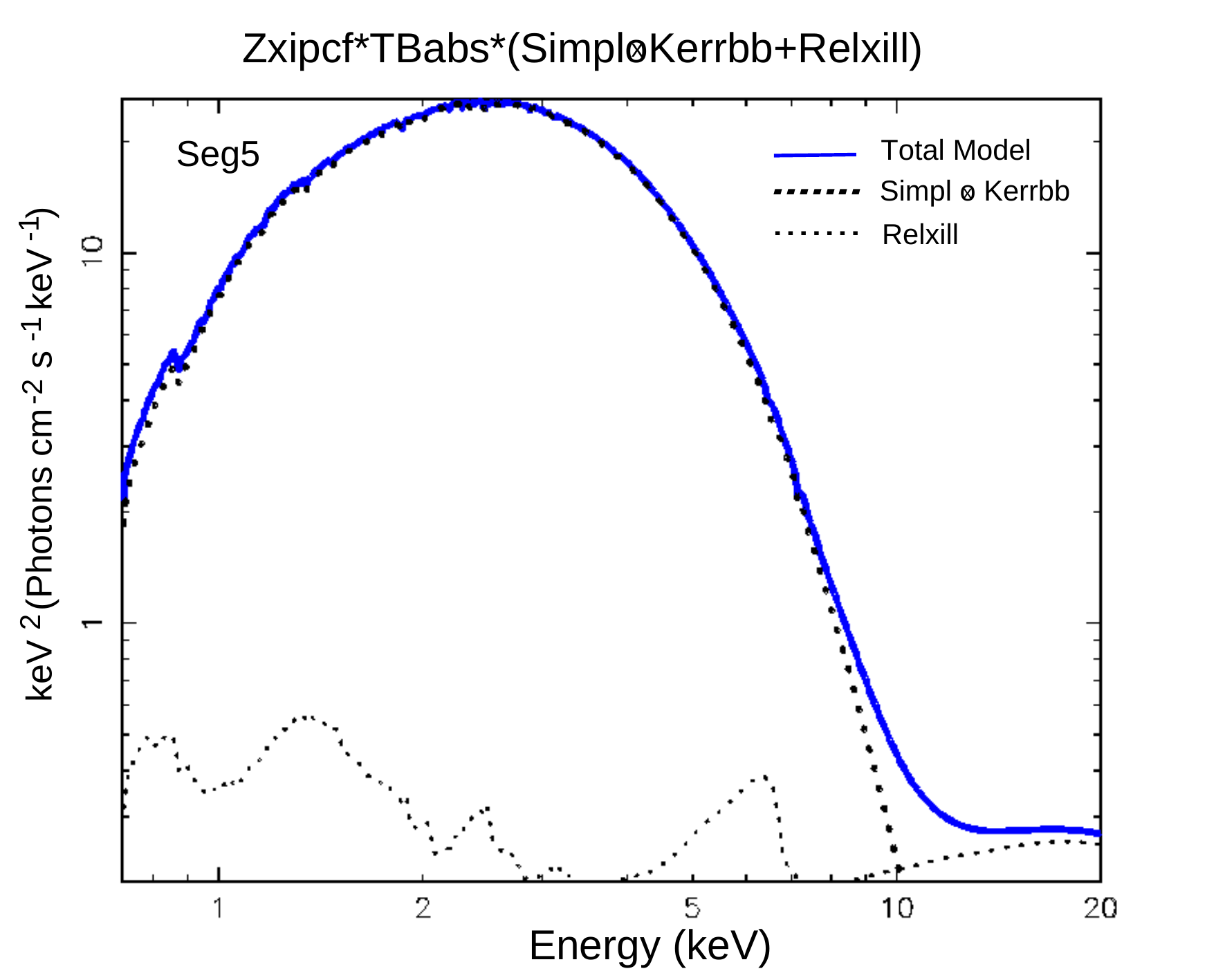}
    \hspace{-5cm}

\caption{Spectral decomposition of the best fit models. Left panel is for Model1 (\textit{Zxipcf*TBabs(Simpl$\otimes$Diskbb+Relxill)}), Middle panel is for Model2 (\textit{Zxipcf*TBabs(Simpl$\otimes$Kerrd+Relxill)}) and Right panel is for Model3 (\textit{Zxipcf*TBabs(Simpl$\otimes$Kerrbb+Relxill)}).The total model contribution is depicted as blue solid line, the reflection component is shown in dotted line and emission from the mildly comptonised disk as dashed line. Both dotted and dashed components include contributions from \textit{Zxipcf*TBabs}.}
\label{fig:spec_decomp}
    
\end{figure*}

The \textit{NuSTAR} epochs of \citetalias{nustar_paper2022} are taken around the time of AstroSat observations, thus following \citetalias{nustar_paper2022},  we next include a warm absorber \textit{Zxipcf} \citep{zxipcf_kallman,zxipcf_reeves}. This model considers a partial absorption characterized by covering fraction (cf) of incoming photons by an ionized absorber whose ionization is described by $log(\xi)$ (we fix it at 2, \citetalias{nustar_paper2022}). The model \textit{Zxipcf*TBabs*(Simpl$\otimes$Diskbb+Relxill)} (Model1 from here on) showed drastic improvement in the fitting, we show the residuals in right panel of Figure~\ref{fig:delchi_dbb_kerrd} along with reduced $\chi^2$ and the best fit parameters in Table~\ref{tab:diskbb}; upper panel. The errors quoted are calculated in the 90$\%$ confidence region. We also show the relative contribution of the model components to the spectrum in Figure~\ref{fig:spec_decomp}; left panel. Photon index takes up a value $>$2.5 for each segment with inner disk temperature remaining in range 0.78--0.85. We observed a clear variation in the $N_{Disk}$ across Segments from Seg4 to Seg8. The inferred estimates of inner radius from the $N_{Disk}$ are reported in Table~\ref{tab:diskbb} with and without considering a spectral hardening ($f_{col}$) of 1.7 \citep{shimura1995spectral}, the values suggest possible truncation of inner disk radius. 

We replace \textit{Diskbb} with alternative disk component \textit{Ezdiskbb} \citep{ezdiskbb} in Model1, the difference is the assumption of a zero-torque boundary condition at the inner edge in \textit{Ezdiskbb}. It also has two parameters, $kT_{in}$ and a normalization given by $(1/f_{col}^4) (R_{in}/D)^2 cosi$ factor. We fit this model to Seg4 to Seg8 and although no significant improvements were observed in the fitting, lower estimates of the normalization and consequently of inner radius were obtained as expected \citep{ezdiskbb}. Despite the lower estimates, the disk was found to be truncated at radii greater than 6~$R_g$ for all segments, considering $f_{col}$ to be 1.7.

Now, we investigate if the changes in spectral hardening could be responsible for the observed evolution of the disk. To do this, we substituted the \textit{Diskbb} component in Model1 with \textit{Diskpn} \citep{diskpn}, which along with a zero-torque boundary condition also has inner disk radius as a fitting parameter and a normalization, 	$M^2\cos(i)/(D^2*f_{col}^4)$. Thus, allowing for independent calculation of $f_{col}$. By setting the inner radius to the minimum acceptable value of 6~$R_g$ in the model and fitting it we obtained results that were equally good as those obtained using \textit{Diskbb}. Therefore, using the normalization values we determined that $f_{col}$ varied in the range of 1.65-1.92 throughout the epochs for a constant radii. The lowest

\FloatBarrier

\begin{figure*}
\hspace{-3cm}
\includegraphics[width=7cm, height=8cm]{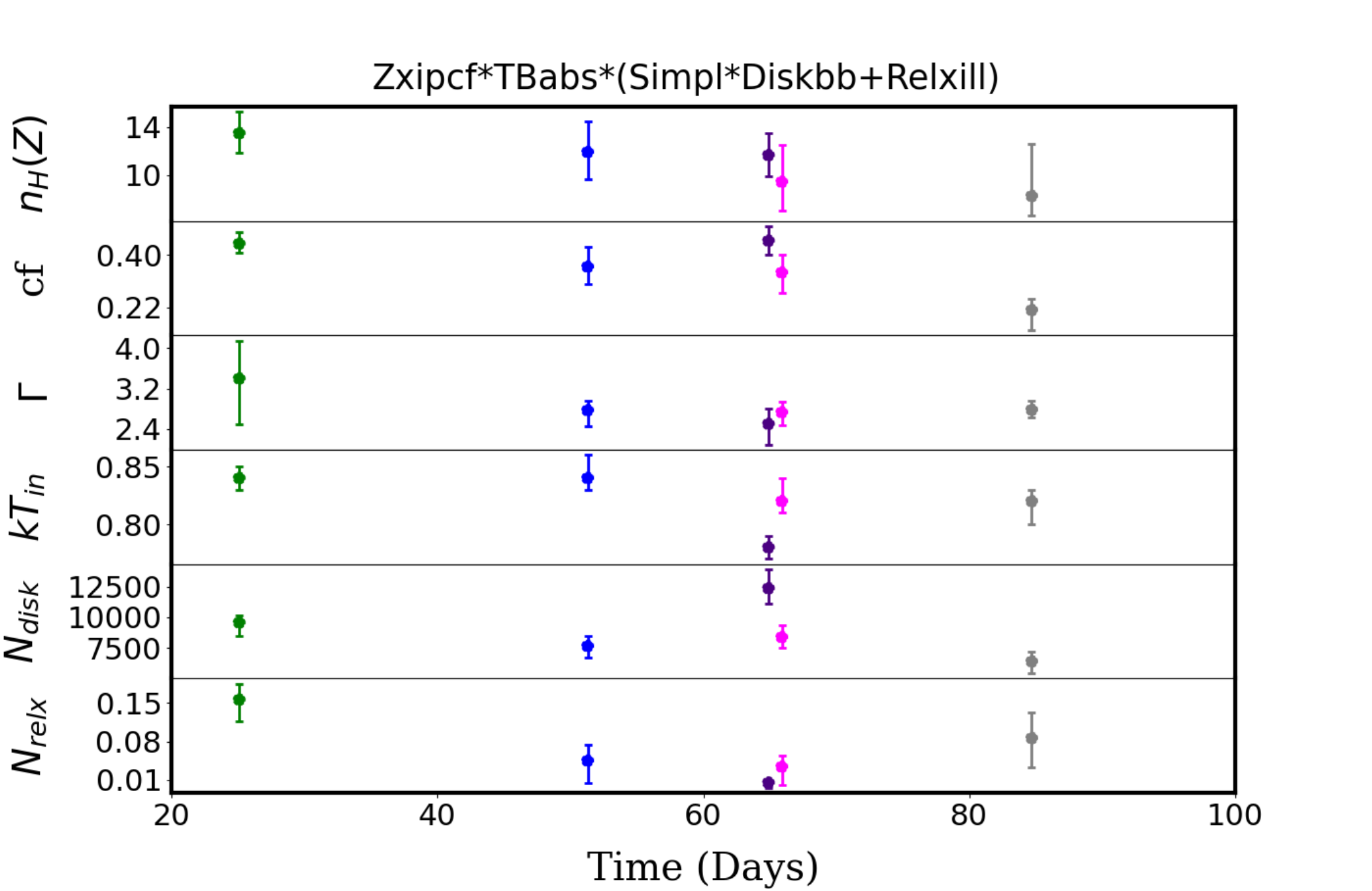}
\hspace{-0.5cm}
\includegraphics[width=7cm, height=8cm]{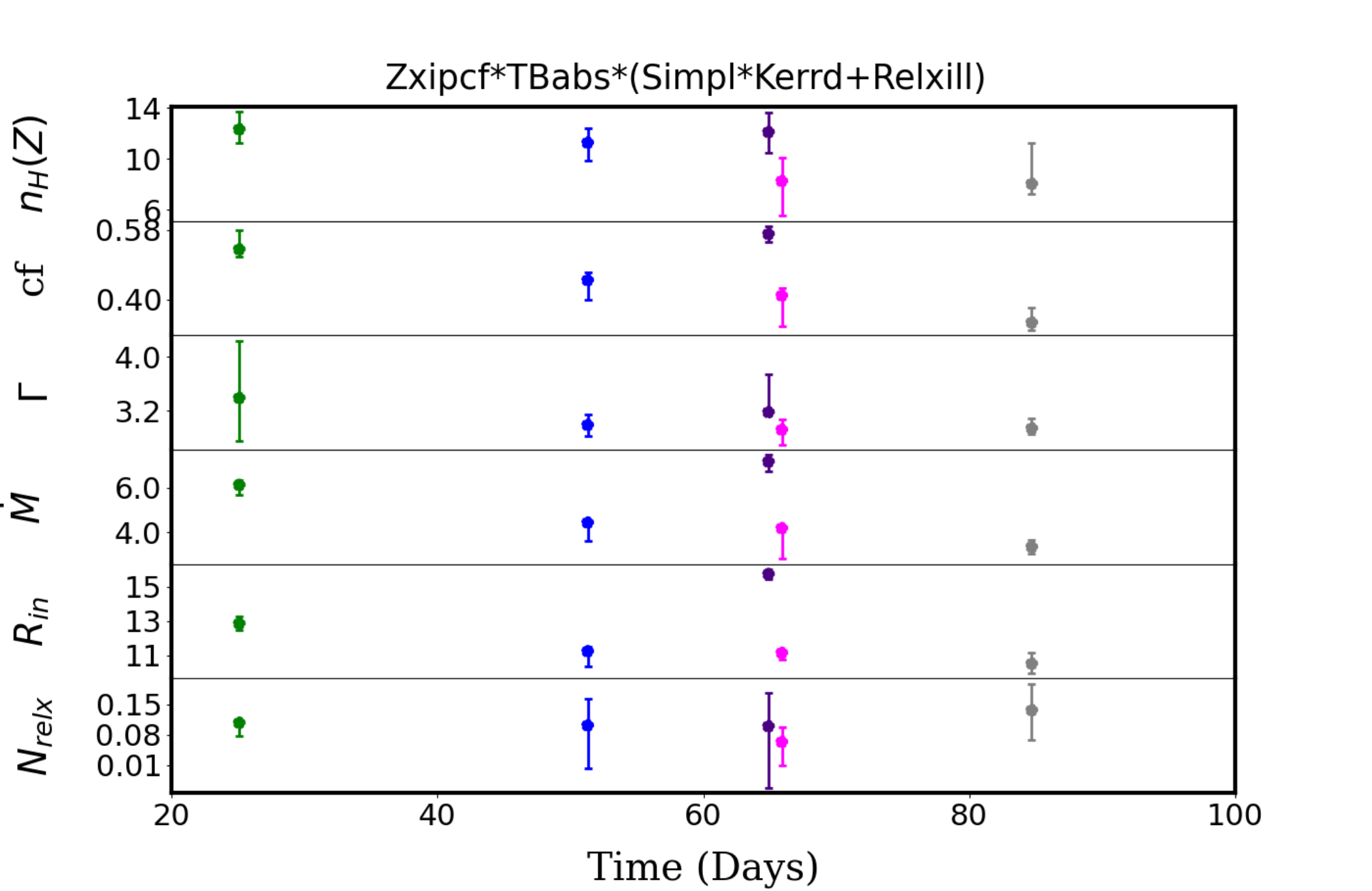}
\hspace{-0.5cm}
\includegraphics[width=7cm, height=8cm]{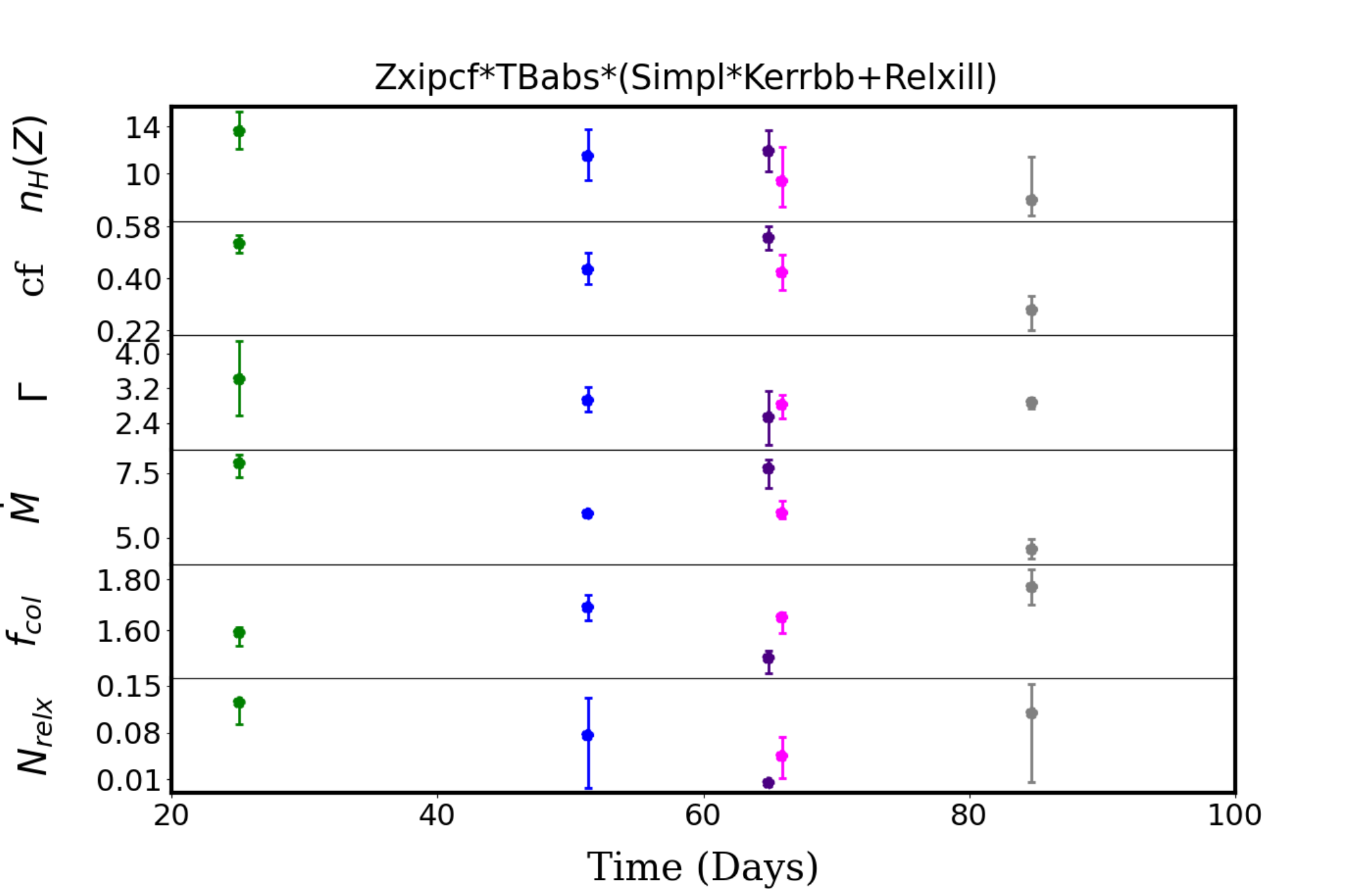}
\hspace{-3.0cm}
\caption{Evolution of spectral parameters for Model1 (left panel), Model2 (middle panel) and Model3 (right panel) for Seg4 to Seg8. Here, $n_H$(Z) is the column density of \textit{Zxipcf} in units $10^{22} cm^{-2}$, cf is the covering fraction, $kT_{in}$ is in keVs, $\dot{M}$ is in units $10^{18} g/s$ and $R_{in}$ in $R_g$. The x-axis mentions the number of days passed since the start of first observation of \bh\ with AstroSat i.e. Epoch1 (59396.11) 
}
\label{fig:evolution}
\end{figure*}

\noindent value of  1.65 was observed for Seg6, while the highest value of 1.92 was found for Seg8.

We know that for a spinning black hole, relativistic effects are bound to play a role in emission especially near to the ISCO, therefore we investigated the inclusion of relativistic disk model \textit{Kerrd} \citep{ebisawa2003} in Model1 replacing \textit{Diskbb}. In addition to both mass of the black hole and accretion rate, this model component allows the inner radius to be treated as a fit parameter, which is important to test the disk truncation results from \textit{Diskbb}. \textit{Kerrd} considers a spinning black hole with spin set to 0.998 for which ISCO will be at 1.235~$R_g$ and thus assumes extreme relativistic effects. A maximal-spin of 0.998 for \textit{Relxill} agrees with the assumption of \textit{Kerrd}. Fixing all other parameters as done for Model1, we fit, what we now refer to as Model2, \textit{Zxipcf*TBabs*(Simpl$\otimes$Kerrd+Relxill)} to spectra on Seg4 to Seg8. Model2 fits the spectra well, we show its best-fit parameters in middle panel of Table~\ref{tab:diskbb}. The relative contribution of the model components to the spectrum is shown in Figure~\ref{fig:spec_decomp}; Middle panel. We find the photon index is consistent to that found for Model1 within error bars and the inner radius is truncated at $>10$~$R_g$ throughout all segments.

Our findings from Model1 suggested that the disk is truncated, and the estimates obtained from the \textit{Kerrd} model produce similar results. However, given that \textit{Kerrd} model assumes a spin of 0.998, we investigate whether a spin of 0.43 \citep{morningstar2014spin} could still produce a good fit. Thus, we replaced \textit{Kerrd} with another relativistic disk \textit{Kerrbb} \citep{kerrbb}, we refer to this model \textit{Zxipcf*TBabs*(Simpl$\otimes$Kerrbb+Relxill)} as Model3. The relative contribution of the model components to the spectrum is shown in Figure~\ref{fig:spec_decomp}; Right panel. For \textit{Kerrbb} we fixed spin to 0.43, $\eta$ to zero (zero torque at the inner boundary), spectral hardening  to 1.7 and normalization to unity by only allowing $M_{BH}$ and accretion rate to vary in the fit. We employed this model to fit the spectra and observed that this fit yielded different mass estimates for all segments, even within a 2 sigma error margin. As next step, keeping the mass constant at 9.4~$M_{\odot}$, we set $f_{col}$ to be free. This enabled us to achieve fits of comparable quality as with Model1 and Model2, with $f_{col}$ displaying a variation in the range of 1.5-1.8. The best-fit parameters are shown in Table \ref{tab:diskbb}. The fitting suggests that when assuming a constant radius for the inner edge, as the assumption holds in the \textit{Kerrbb} model, it is necessary to consider the variation of $f_{col}$ to explain the spectral variation of the source.

Finally, we explore the possibility of disk deviating from the standard hypothesis \citep{shakura1973} by implementing another multi-temperature disk emission model \textit{Diskpbb} \citep{diskpbb} in place of \textit{Kerrbb}. This model introduces exponent (p) of the radial temperature dependence ($T_{in} \propto R^{-p}$) along with $T_{in}$ and normalization, $(R_{in}/D_{10})^2 cos(i)$ as fit parameters. In the standard disk model the value of `p' is 0.75, deviations from this value may indicate departure from standard assumption. Initially, we fit the model to the Seg4 to Seg8 allowing both `p' and normalization to vary. The fitted results were similar to those obtained using Model1 and Model2, except that the best fit value of p turned out to be  $>$0.75, which is in contradiction to $p \sim 0.5$ appropriate for a slim-disk model \citep{slimdisk}. Even when the normalization was fixed to a value corresponding to the $R_{in}$ being at the ISCO, the best value of p remained $>$0.75. Finally, fixing $p=0.5$ yielded larger  $\chi^2$  with $\Delta \chi^2$ $>$ 40, as compared to Models 1, 2 and 3.

Therefore following our spectral analysis and fit estimates, we only test Model1, Model2 and Model3 for off-axis segments with the LAXPC spectra. For all three models we also examined the changes in their various spectral parameters observed with time over from Seg4 to Seg8, this is shown in Figure~\ref{fig:evolution} (Model1:left panel, Model2:middle panel and Model3:right panel). Specifically, the parameters are:  $n_H(Z)$ (of \textit{Zxipcf}), covering fraction (cf), photon index ($\Gamma$), inner disk temperature ($kT_{in}$), \textit{Diskbb} normalization ($N_{Disk}$), \textit{Relxill} normalization ($N_{relx}$), accretion rate ($\dot{M}$), inner radius ($R_{in}$) and spectral hardening ($f_{col}$). We observed a decreasing trend in the absorption parameters ($n_H(Z)$ and cf) along most segments, with the exception of a slight increase in Seg6. Similar changes were noticed in the $\dot{M}$ and $R_{in}$. Additionally, the $f_{col}$ exhibited variations over the course of our on-axis observations.

\subsection{Off-axis observations}

Due to a significant offset angle with the LAXPC, the source was located outside the field of view of the SXT instrument for Seg1 to Seg3. As a result, we only analyzed the data from the LAXPC instrument. We applied Model1, Model2 and Model3 while fixing the values of parameters corresponding to the components active in soft X-ray regime, mainly $n_H$ of \textit{Zxipcf}, the covering fraction, and $n_H$ of \textit{TBabs} to the values obtained for Seg4, which is the closest observation to the off-axis segments. The spectra were effectively fitted by all models, and their respective best fit parameters are presented in Table~\ref{tab:diskbb}. Model1 produced significantly high values of $N_{Disk}$ for all three segments, and the inner disk temperature was $\sim$ 0.93~keV for Seg1, which was the highest among all segments. These high estimates of $N_{Disk}$ indicate an increased truncation during the high flux states. Model2 revealed a super-Eddington flow with accretion rate ranging from $\sim$15--17$\times$$10^{18} g/s$ for the three segments, and a truncated disk radius ranging from $\sim$15--19 $R_g$. Whereas Model3 fitting shows that $f_{col}$ had to change from $\sim$1.27-1.42 in order to explain the spectral evolution seen from Seg1 to Seg3.

\section{Discussion}

After entering the outburst, \bh\ rapidly attained its peak flux within a few days, leading to its observation primarily during the decay phase by multiple missions, including \textit{AstroSat}. We divided the nine observations taken with \textit{AstroSat} into eight segments based on the hardness ratio of 6.0--20.0/4.0--20.0 keV, as depicted in Figure \ref{fig:maxi} (right panel). Throughout all these segments, the source consistently remained in a High/Soft state, where the dominant emission originated from a thermal disk, contributing to over 95$\%$ of the total flux. Based on the analysis presented in Section \ref{sec:Analysis}, we have selected three final models: Model1 (Diskbb), Model2 (Kerrd), and Model3 (Kerrbb), which describe the data well. Utilizing Model1, we observed a relatively high inner disk temperature ranging from approximately 0.78 to 0.85 keV, which exhibited consistency across most segments. However, in Seg1, a slightly elevated temperature of around 0.93 keV was detected. The degree of comptonization was found to be weak, with only a small fraction of photons interacting with the hot electrons. In order to capture the spectral evolution seen during the observations, we employed various disk models which have been discussed in Section \ref{sec:Analysis}. The interpretation of our findings is further discussed in the subsequent sections.

\subsection{Truncation of the accretion disk in soft state}

We found large disk normalization with \textit{Diskbb}, therefore, we derived the spectral hardening corrected inner disk radii from these values  which are found to be truncated at radii $>$13~$R_g$, with significantly higher truncation for the off-axis segments with $R_{in}$ reaching up to 24~$R_g$. Additionally, without spectral hardening the estimated inner radius is lower but still higher than the ISCO for a black hole spinning at 0.43 whose ISCO is at 4.52~$R_g$ (see the entry for $R_{in}^{'}$ in upper panel of Table~\ref{tab:diskbb}). The estimates without the spectral hardening are consistent with the ones reported in \cite{park2004}, which have attributed the lack of spectral hardening to the lack of disk atmosphere in HSS. Even the estimates from \textit{Kerrd}, which incorporates strong relativistic effects gives a large truncated radius with $R_{in}$ $>$11~$R_g$ for all segments (and a larger truncation for the off-axis segments with $R_{in}$ $>$15~$R_g$). We should also mention that fixing $n_H$ (\textit{TBabs}) to 0.5 $\times$ $10^{22}$ $cm^{-2}$ does not alter our results of truncation.  It should also be noted that since the distance estimates are well-constrained for this source, the inner radius can be reliably estimated from the modelling.

The common consensus for a BHB in High/Soft state is that the disk is at a constant radius which is inferred as the ISCO \citep{gierlinski2004black, steiner2010constant}. However, our findings using two different disk models, \textit{Diskbb} and \textit{Kerrd}, indicate that the disk is not constant and exhibits signs of truncation. The large normalization values of \textit{Diskbb} are consistent with values reported in \citetalias{nustar_paper2022} and \cite{prabhakar2023} in their study of \bh. During low Eddington fractions, the truncation of the disk may not be due to evaporation of disk and formation of corona but rather due to mass loss from the disk wind \citep{taka2006}. While we have a good understanding of the accretion flow geometry at low mass accretion rates \citep{done2007}, our comprehension is inadequate regarding the involvement of strong outflows such as disk winds at high accretion rates.

\begin{figure}
    \centerline{
    \includegraphics[trim=0cm 0cm 0cm 2.93cm, clip=true, width=8.8cm, height=5.0cm]{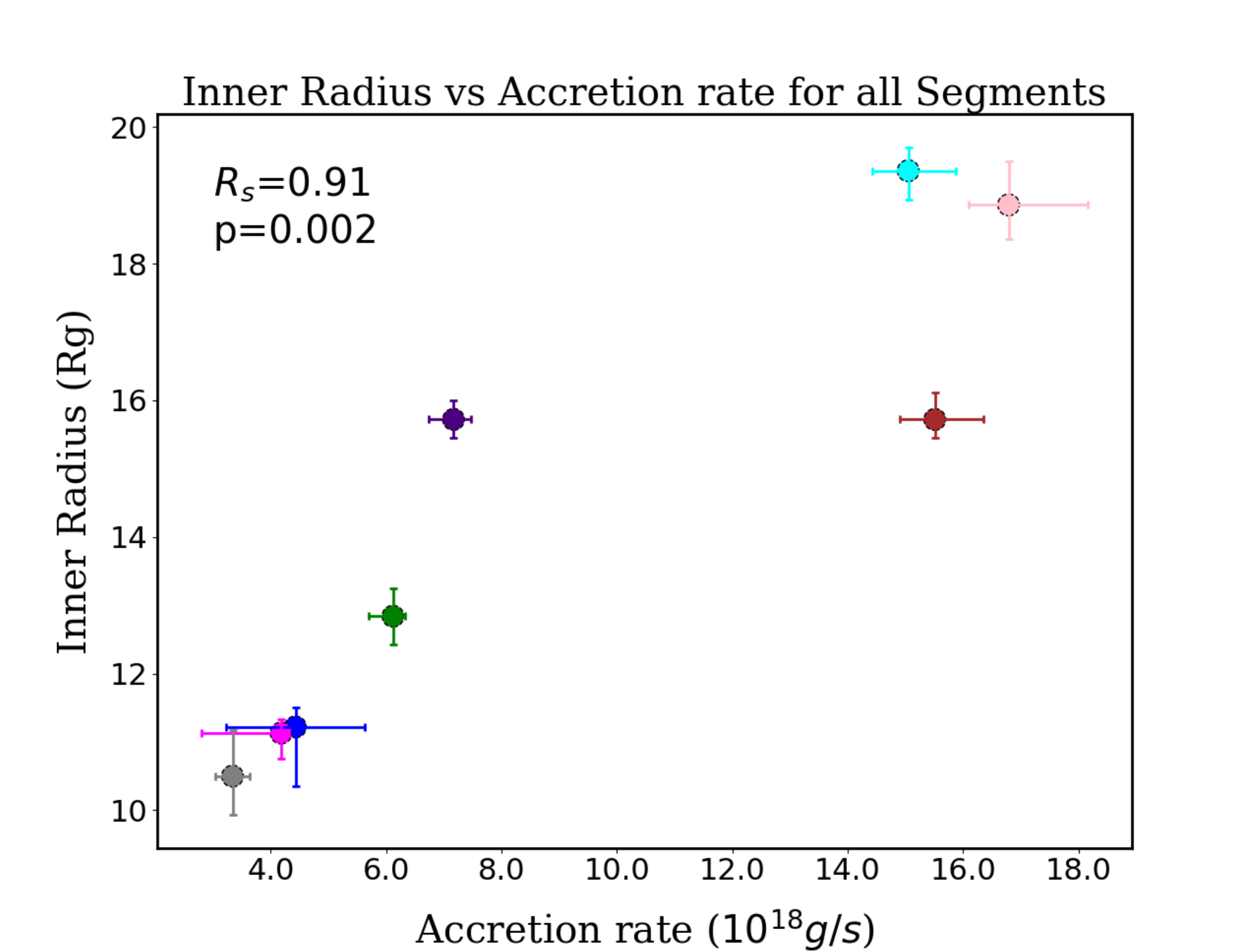}}
    \caption{Correlation between Inner disk radius ($R_g$) and accretion rate ($10^{18} g/s$) for all segments. The plot follows the color scheme of Figure~\ref{fig:maxi}. The three points in the upper right of the panel correspond to the off-axis segment. The high value of Spearman rank coefficient ($R_s$=0.91) and the low chance probability (p-value=0.002) indicates a strong correlation between the accretion rate and the inner disk radius in the HSS. 
    }
    \label{fig:correlation_kerrd}
\end{figure}

For a 9.4$M_{\odot}$ black hole the Eddington luminosity\footnote{Eddington luminosity is calculated by $L_{Edd}$ $\approx$1.26$\times$$10^{38}$$M/M_{\odot}$ ergs/s} is 1.18 $\times$ $10^{39}$ ergs/s. From this, one can determine the critical accretion rate $\dot{M}_{Edd}$=$L_{Edd}/c^2 \eta$, where $\eta$ is the accretion efficiency of the disk. By taking $\eta$=0.1 for thin disk, we have $\dot{M}_{Edd}$=13.2 $\times$ $10^{18}$ g/s. Comparing the accretion rate estimated using Model2 to the critical accretion rate, we find that accretion rate is at sub-Eddington values for the on-axis segments and at super-Eddington for the off-axis segments. This signifies that the geometrically thin disk assumption may not be valid for the off-axis segments and instead more complicated models are required given the possibility of puffed-up disk due to acting radiation pressure. For Segments 4 to 8, we have observed a simultaneous decrease in both the accretion rate ($\dot{M}$) and the inner disk radius ($R_{in}$). To explore the potential correlation between these two variables, we have generated a plot of $R_{in}$ versus $\dot{M}$ in Figure \ref{fig:correlation_kerrd}, considering both on- and off-axis segments. By employing the Spearman rank correlation coefficient, we have successfully quantified their relationship, revealing a robust correlation with a rank coefficient ($R_s$) of 0.91 and a p-value of 0.002.

\citet{mudambi2022spectral}, conducted a similar study on 4U~1957+115 where a positive correlation between accretion rate and inner disk radii or equivalently between the accretion rate and spectral hardening similar to the one reported here was found. Both 4U~1957+115 and \bh\ were found in the High/Soft state with the disk flux $>85$\% of the total flux. It must be noted that the results reported by \citet{mudambi2022spectral} depicts the evolution of the accretion disk structure of 4U~1957+115 for over a period of 3 years whereas, we see similar evolution in the span of a few weeks. Both of these variations are in contrast with the typical behaviour of BHBs when they transition from LHS to HSS, where over the time-scales of days to weeks, the accretion rate is typically inversely correlated with inner disk radius. \cite{garg2022} also found a positive correlation between the two quantities for another transient MAXI J1535-571, although it should be mentioned that the source was in Hard-Intermediate state. Meanwhile, the spectral analysis of short-term variability ($\sim$minutes) of GRS 1915+105 by \cite{rawat2022}, have revealed a direct correlation of the inner disk radius with the accretion rate but at shorter timescales, \citet{bhargava2022MNRAS.512.6067B} found for Cyg X-1 that inner disk variation may be independent of the accretion rate. 

\subsection{Alternative interpretation for soft-state evolution}

We explored alternative explanations to the varying truncated disc interpretation. Specifically, changes due to spectral hardening ($f_{col}$) were considered. It was noted that the disk atmosphere can comptonize the thermal photons emanating from the disk, resulting in spectral hardening. Since $f_{col}$ can vary with the changing accretion rate or the coronal power \citep{davis2005}, and the estimation of the disk's inner radius using the \textit{Diskbb} model relies on this parameter, it was important to account for its potential variation. To investigate this, two models, \textit{Diskpn} and \textit{Kerrbb}, were employed. With the \textit{Diskpn} model, it was found that $f_{col}$ ranged from 1.6 to 1.9, while the inner radius was fixed at 6$R_g$. The \textit{Kerrbb} model, which assumes the inner radius at the ISCO, failed to fit the data (\citetalias{nustar_paper2022}, \cite{prabhakar2023}) without allowing $f_{col}$ to vary. The observed variation of $f_{col}$ ranged from 1.5 to 1.9. These values derived from both models are consistent with the boundaries $1.4<f_{col}<3.0$ obtained from theoretical predictions \citep{shimura1995spectral,merloni2000}.

It is important to highlight that the source achieves a significant fraction of its luminosity, 0.2~$L_{Edd}$ during on-axis segments, and upto 0.6~$L_{Edd}$ during off-axis segments. Therefore, another possibility has to be considered that the disk may not be a standard disk \citep{shakura1973}, due to involvement of other factors like strong radiation pressure. At such high luminosities, certain sources (GRO J1655–40, XTE J1550–564 see \cite{done2007}) have
exhibited deviations from the standard relation of $L_{Edd}$$\propto$$T_{in}^{-3/4}$. The existence of wind-like strong absorption also imply that radiation pressure might contribute to altering the thin structure of the disk. This phenomenon aligns with the model proposed by \cite{taka2006}, where radiation pressure may inflate the inner region of the disk, causing it to recede and subsequent truncation. Even though \textit{Diskpbb} model suggests radial temperature dependence exponent to be $>0.75$, but better data statistics are required to test these assumptions further.

\subsection{Nature of the additional absorption}

There are also clear indications of blurred reflection in the spectra particularly the prominent dip observed in the energy range of 7.0--11.0~keV but no clear indication of the Fe k$\alpha$ emission line especially in the SXT data. This is probably due to limited spectral response at such energies ($\geq4.0~keV$), making it challenging to model the spectra using only relativistic line models. We tried to account for absorption feature using different reflection models, however, we found the best fit when \textit{Relxill} is used with a combination of model \textit{Zxipcf} following methodology of \citetalias{nustar_paper2022}. Covering fraction values are comparable to those found in \citetalias{nustar_paper2022}, although they find higher values of absorption column density.  Signature of disk wind includes detection of highly ionized absorption lines; Fe XXV and Fe XXVI at $6.96$~keV and $6.7$~keV respectively, therefore it could have been possible that the lines were present but could not be resolved given low spectral resolution of instruments. This could be properly checked by utilizing data from Chandra observatory which provides high resolution grating spectra to resolve such narrow features, although during our observations the extreme brightness of source would induce huge pile-up effects in Chandra spectra. 

\begin{figure}
    \centerline{
    \includegraphics[trim=0cm 0cm 0cm 2.5cm, clip=true, width=9.0cm, height=5.5cm]{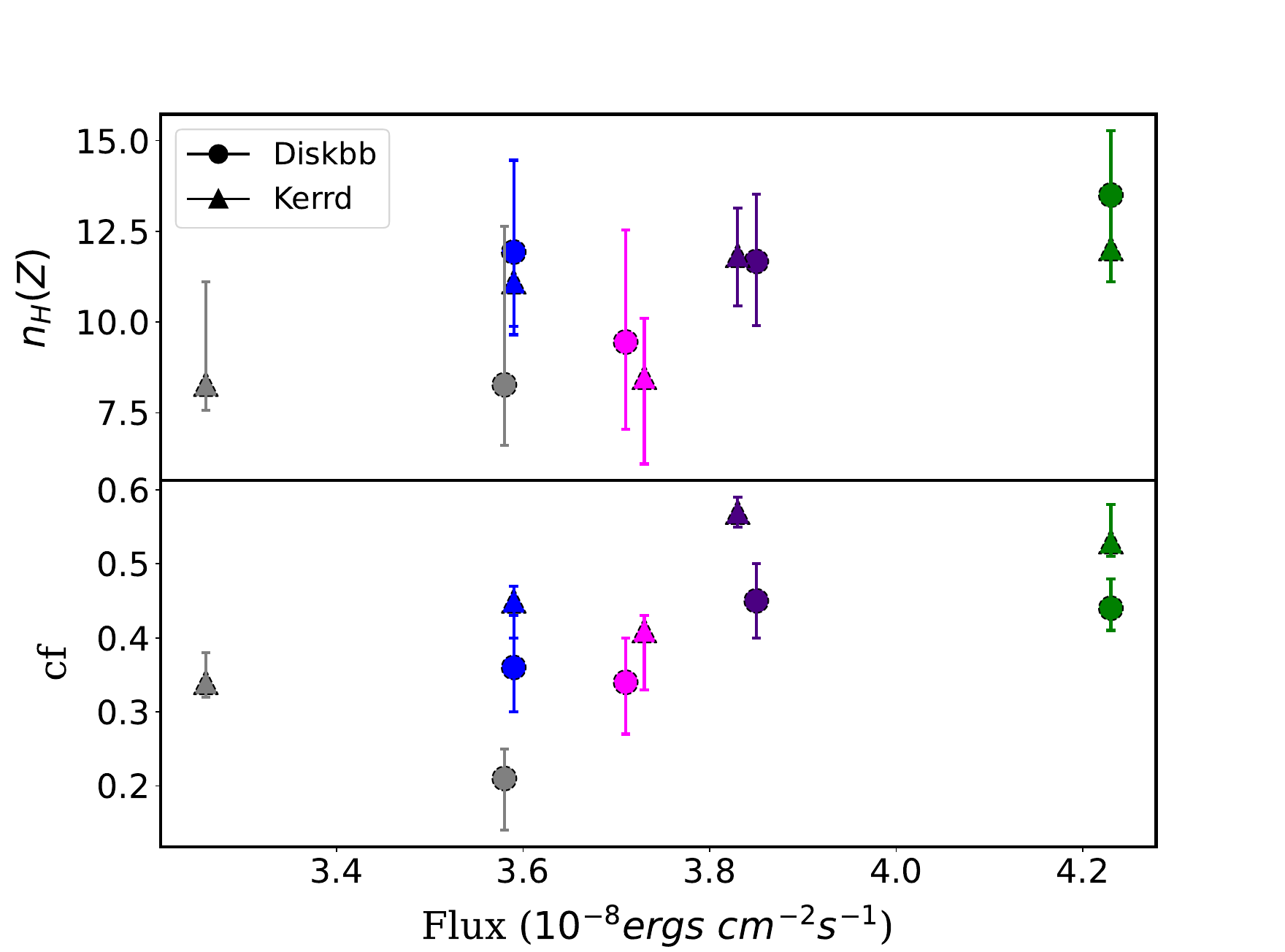}}
    \caption{Evolution of column density ($n_H(Z)$ $10^{22}~cm^{-2}$) and covering fraction (cf) with absorbed flux in 0.7-20.0~keV ($10^{-8}~ergs~s^{-1}~cm^{-2}$) for on-axis segments. Evolution is shown for Model1 (circles) and Model2 (triangles). The plot follows the color scheme of Figure~\ref{fig:maxi}.}
    \label{fig:evolution_zxipcf}
\end{figure}

As illustrated in Figure~\ref{fig:evolution}, we find that there is a definite decrease in both absorption column density ($n_H(Z)$) and covering fraction (cf) as the source moved along the decay phase, showing external absorption (possibly ionized wind) got weaker by the last observation. As for $\Gamma$ it is more or less consistent within errorbars for all segments. We also plot the two parameters $n_H(Z)$ and cf against the absorbed flux for the on-axis segments as shown in Figure~\ref{fig:evolution_zxipcf}, for Model1 and Model2. $n_H(Z)$ and cf are both seen to increase as the flux increases, suggesting the absorption is more prominent in the higher flux states. This correlation aligns with the findings of \cite{prabhakar2023}, who reported decrease in the strength of absorption after MJD 59430 with the decreasing flux. Observations taken before MJD 59430 were found to show opposite behavior where the strength increased with the decreasing flux. Since, AstroSat's on-axis epochs were taken close to MJD 59430 therefore we only study the period where the strength decreases in correlation with flux. Here, we associate the strength of absorption to \textit{Zxipcf's} absorption column density and particularly to covering fraction. Evolution of such wind-like features with flux have been earlier reported in \cite{trigo2007, kubota2007, done2007}

\section{Conclusion}

{During its exceptionally bright outburst in 2021, \bh\ emerged as one of the brightest transients ever observed. Our analysis of AstroSat data during the decay phase indicates that the spectral evolution in the soft state necessitates either a notable truncation of the inner radius or variations in the spectral hardening factor, as demonstrated by different disk models. If we consider truncation, this departure from the usual behavior where the disk is typically situated at the ISCO regardless of luminosity represents a distinctive discovery. Moreover, by employing a partially ionized absorber, we successfully modeled a prominent absorption dip observed in the spectra. This absorption feature exhibited an evolution that closely correlated with the flux of the observations, thus resembling with the characteristics of a wind-like outflow. Unfortunately, AstroSat did not capture the source during its hard state, which would have provided additional insights into the evolution of the absorption feature during a significant decrease in disk contribution, as well as the study of timing characteristics in that state.}

\section{Acknowledgements}

We would like to thank the anonymous referee for their valuable suggestions, which improved the quality of this work. This work has utilized data from \textit{AstroSat} mission which is archived at Indian Space Science Data Centre (ISSDC). We are grateful to the SXT and LAXPC POC teams for providing the data and requisite softwares to perform data analysis. NH acknowledges the financial support provided by Department of Science and Technology (DST) under the INSPIRE fellowship scheme. AG, RM and SS acknowledges the financial support provided by Department of Space, Govt of India (No.DS$\_$2B-13012(2)/2/2022-Sec.2). 

\section{Data Availability}
The data used in the publication is publicly available for download at \url{https://astrobrowse.issdc.gov.in/astro_archive/archive/Home.jsp} using the observation IDs mentioned in Tab~\ref{tab:log}. 

\bibliography{main} 

\end{document}